\documentclass[conference, onecolumn, draftcls]{IEEEtran}
\usepackage{authblk}
\usepackage{amsmath,amssymb,amsfonts}
\usepackage{algorithm}
\usepackage{algorithmic}
\usepackage{listings}
\usepackage{hyperref}
\usepackage{graphicx}
\usepackage{textcomp}
\usepackage{stfloats}
\usepackage[numbers,sort&compress]{natbib}
\usepackage{float}
\usepackage{xcolor}
\def\BibTeX{{\rm B\kern-.05em{\sc i\kern-.025em b}\kern-.08em
    T\kern-.1667em\lower.7ex\hbox{E}\kern-.125emX}}

\makeatletter
\def\ps@IEEEtitlepagestyle{%
  \def\@oddhead{}% remove header
  \def\@evenhead{}%
  \def\@oddfoot{\hfil\thepage\hfil}% page number only
  \def\@evenfoot{\@oddfoot}%
}
\makeatother

\begin{document}

\title{Efficient Parallel Compilation and Profiling of Quantum Circuits at Large Scales}

\author[1,*]{Jane Moore}
\author[1]{Michael Hart}
\author[1]{John McAllister}

\affil[1]{Queen's University Belfast, Belfast, United Kingdom}

\maketitle
\thispagestyle{IEEEtitlepagestyle}

\begin{abstract}
Compiling quantum circuits is a major bottleneck in quantum computing, and given the scale required in a few years, is likely to become infeasibly long. Techniques to reduce compilation time for quantum circuits are sorely needed. Furthermore, resources to test acceleration techniques are similarly lacking due to the limited scale of circuits in benchmark suites and mismatches in characteristics of these circuits and those produced by random circuit generators. This paper resolves the latter of these problems by describing a random circuit generator which allows control of circuit density, width and depth parameters. This is used to derive 8000 experimental large-scale circuits and test a novel approach to compiler parallelisation. This separates a circuit into sub-circuits which are compiled in parallel and recombined to produce a compiled circuit. When the parallel approach was tested using Qiskit, a peak speedup of 15.56 was achieved with corresponding overheads of less than 1\%.
\end{abstract}

\section{Introduction}
Quantum circuits describe sequences of quantum logic gates operating on qubits. Like classical computer programs, they need to be compiled for a target device before they can be executed on that device \cite{correct_compilation} \cite{Venturelli_2018}. Compilation transforms the circuit into a series of basic gate instructions in a language such as QASM, accounting for target-specific constraints, such as the viable gate set of the processor and the topology of its qubit connectivity network, as well as optimisations for more efficient execution \cite{QAOA_compilation}.

Compilation is a bottleneck in the current quantum computing pipeline \cite{wille_compilation_time}. A circuit of 100,000 gates deep and 200 qubits takes over 24 hours to compile with current compilers; previous work has shown that compilation time surpasses simulation time for circuits including more than 100 qubits, implying it scales even more aggressively with circuit size than even the notoriously intractable problem of quantum simulation \cite{Nation2025}. Given that current trends in quantum algorithm design suggest millions of qubits will be required in the future \cite{factor2048} \cite{PhysRevA.70.052328} \cite{gambetta_future}, there is a keen need for methods for accelerating quantum compilation.

Current work in this area is somewhat constrained. A diverse range of factors influence compilation time, including number of qubits, number of gates and circuit density. But current works \cite{parallel_paper}\cite{wille_compilation_time} are all restricted in at least one of these factors due to a more general problem - difficulty obtaining test circuits for large-scale, empirical analysis of the quantum compilation process. This is because the number of large-scale circuits in benchmarking suites is very limited \cite{qasmbench} \cite{mqtbench}, with only a handful of circuits with more than 500,000 gates, for instance. Ideally, random circuit generators could address this shortage, but current circuit generators do not provide sufficient control of the properties of the circuits they generate to reflect those of benchmark suites; specifically, whilst number of qubits and gates are controllable, density - a key factor influencing compilation time - is not. The infrastructure to properly test parallel compilation on large-scale circuits is not currently available, hence work to solve the compilation problem is also limited. To the best of our knowledge, there has been no work which considers circuits larger than 100 qubits and 100,000 gates - orders of magnitude smaller than envisaged circuit sizes. This paper resolves both of these problems.

Specifically, a novel random circuit generation technique and generator is presented which allows users to derive circuits which meet density criteria. Given this capability we devise and test a parallelisation technique which is compiler-independent. The effectiveness of this technique is demonstrated across benchmark and randomly-generated circuits across a spectrum of scales (qubits, gates and densities) and compilers. 

Specifically, the following contributions are made:
\begin{enumerate}
    \item To properly profile the performance of the proposed work, a method of generating random circuits of given density is implemented. It is shown how this generator can derive circuits whose density closely match that of existing benchmark suites by enabling generation of 400 random circuits of varying depth, width and gate density characteristics. This supports compiler testing at a level of robustness not previously recorded - up to 200 qubits (twice the previous limit), circuits of more than 16 million gates (previous limit was 100,000).
    \item A compiler-independent method of parallelisation of qubit routing, a key bottleneck operation in quantum compilation, is proposed. By distributing a large circuit across individual cores in a multicore architecture and recombining their compiled versions, it is shown how this approach achieves significant speedup at almost no cost overhead. When tested using Qiskit's SabreSwap and BasicSwap algorithms, peak speedups factors of 12.95 and 15.56 were observed on 16-core devices; these reside alongside minimal increases in circuit gate, SWAP and depth measures of 0.2\%, +0.25\% and +0.85\% respectively. Speedup by a factor 19.8 is observed for TKET.
\end{enumerate}

Existing literature shows how quantum circuit compilation transforms a circuit into a form which can be executed on a target processor. It employs transformations to ensure that the resulting circuit uses only the gate set permitted by the processor, respects the physical qubit layout of the device and applies optimisation techniques, such as gate optimisation to improve the efficiency, robustness or cost of the solution \cite{correct_compilation} \cite{Venturelli_2018} \cite{QAOA_compilation}. 

One particular objective of compilation has prompted a lot of research activity: qubit routing. Whilst in a quantum circuit any combination of qubits can be combined in a single gate, in reality this is not currently generally possible. 
Solid-state quantum processors employ Nearest Neighbor Architecture (NNA) qubit lattices and can only combine qubits when connected in the lattice \cite{NNA} \cite{Muller2024} \cite{solid_state} \cite{comp2arch}. Hence, SWAP operations are inserted during compilation to alter the logical-to-physical qubit mapping to obey NNA \cite{reordering}. The number of SWAP gates in the final circuit is a key compiler quality metric \cite{QAOA_compilation}, along with others such as total gate count and circuit depth \cite{synth_time_optimal}. 

Qubit routing is an NP-Complete problem \cite{Qubit_allocation} \cite{compilation_complexity} and even heuristic solutions can be computationally demanding, which means finding a solution of sufficiently high quality is time consuming. This makes compilation a bottleneck in the quantum computing pipeline \cite{wille_compilation_time}. Since the complexity of these algorithms grow in scale with the size of the circuit \cite{factor2048} \cite{PhysRevA.70.052328} \cite{gambetta_future}, if circuits reach the sizes anticipated in future - up to millions of gates and/or millions of qubits - the compilation times required will be infeasible. Methods to reduce compilation time for large-scale circuits without significantly compromising the quality of results are required.

There has been a lot of work on methods to accelerate compilation. Much has addressed routing algorithms, proposing sub-optimal heuristics which find high-quality logical-to-physical qubit mappings in shorter and shorter periods of time \cite{SABRE_PROP} \cite{Wagner2023} \cite{qmap}. The work in \cite{wille_compilation_time} accelerates compilation when there is existing knowledge of a circuit's structure and target device; it is not apparent how this accelerates the general compilation process for arbitrary circuits. Commercial compilers themselves also offer parallelisation options - for instance, Qiskit \cite{qc_with_qiskit} enables parallelisation of the SabreSwap routing algorithms by allowing several `trials' at once; however each trial is a full repetition of the routing process on the same circuit and hence for very large circuits the same issue of long run-times will re-surface \cite{SABRE_PROP}. The work in \cite{parallel_paper} describes a method of acceleration by parallelising the qubit routing process, but is only tested on relatively small circuits of up to 20 qubits.

In addition to exploring the acceleration of quantum compilation, there has been work done on distributing or parallelising the quantum compilation process. OptQC \cite{OptQC} and Pandora \cite{pandora} consider the parallelisation of quantum compilation, however these methods do not consider hardware-aware optimisations or qubit routing. The work in \cite{ai-partitioning} considers how Large Language Models may be used to partition quantum circuits for compilation and execution. This enables wide circuits to be cut to be compiled and executed on processors with fewer qubits, however there is no exploration of the runtime for this process. Further to this, in the wake of distributed quantum computing, where larger quantum circuits are to distributed over multiple devices for execution \cite{comp_design_for_DQC} \cite{optimised_comp_for_DQC} \cite{modular_compilation_for_DQC}, again this considers cutting the circuit in terms of the circuit width. This requires special consideration of the topology, not just the qubit layout, but also the non-local gates. The scale of the benchmark circuits used in these works are limited by algorithm type or circuit scale.

If compilation for large-scale circuits is to be effectively studied, then it needs to be empirically verified on circuits of representative scale and complexity. In this respect, current work is also limited. The work in \cite{parallel_paper} is verified on a small number of circuits whose size in some respects is orders of magnitude smaller than required for next-generation circuits - it's limited to circuits of 20 qubits and smaller; this is perhaps 10,000 times smaller than the scale of forthcoming circuits. The work in \cite{wille_compilation_time} does study circuits of up to 100 qubits, or 100,000 gates - again, significantly below the scales required - and it studies variants of just two different circuits. 

\begin{figure}[tbp]
    \centering
    \includegraphics[width=0.8\linewidth, trim={0cm 0cm 0cm 0cm},clip]{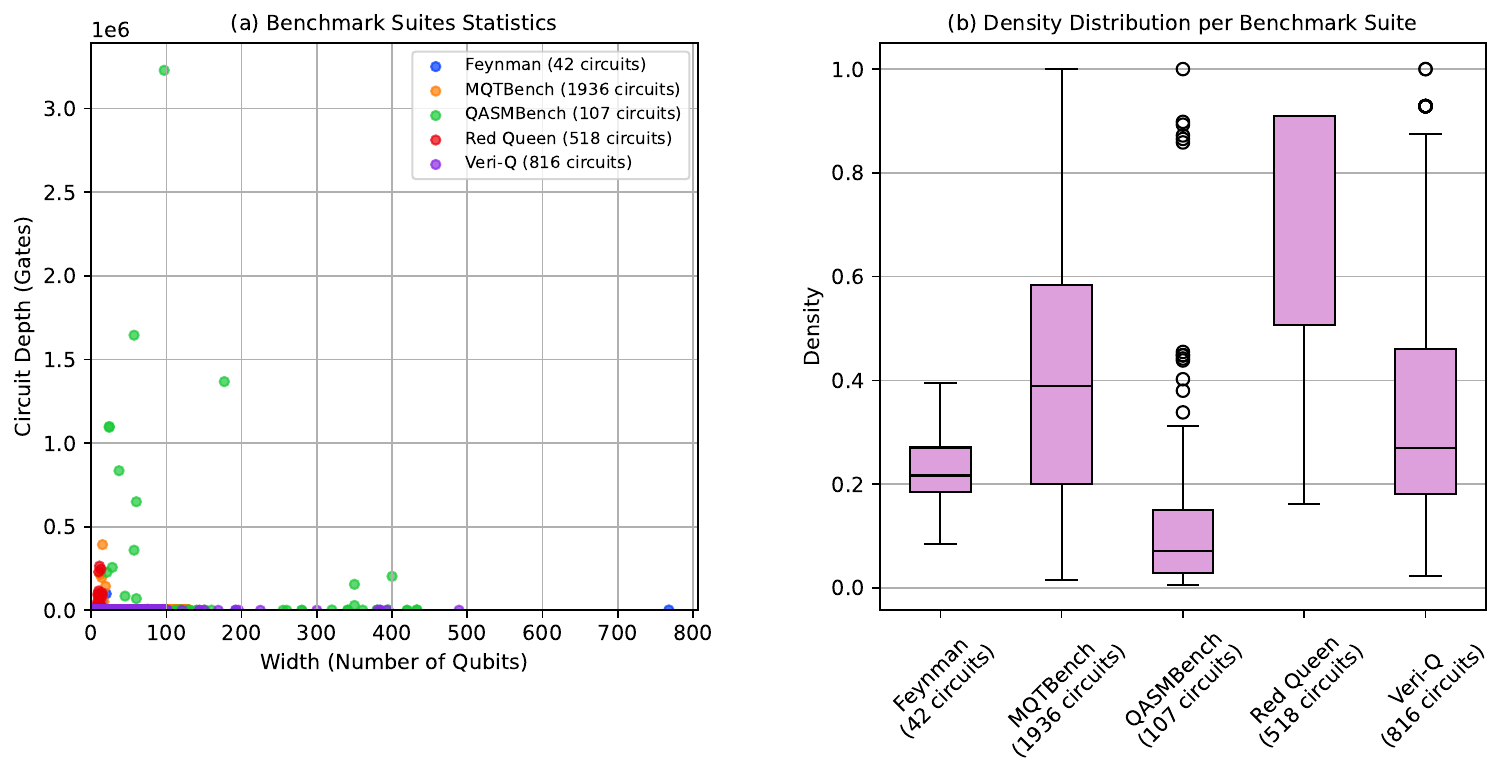}
    \caption{a) Scatter chart illustrating the depth and width statistics for 5 benchmark suites with gate density. b) Box plot illustrating the density distribution of the 5 benchmark suites.}
    \label{fig:combined_gate_depth_denisty_stats}
\end{figure}

Typically, compilers are verified on pre-existing benchmark suites of circuits \cite{qasmbench} \cite{red-queen} \cite{mqtbench} \cite{feynman_suite} \cite{supermarq} \cite{veri_q}. The distribution of circuit size (widths and depths) for these benchmark suites is illustrated in Fig. \ref{fig:combined_gate_depth_denisty_stats}a). 

Red Queen \url{(https://github.com/Qiskit/red-queen)} contains circuits of two general classes - up to 20 qubits and 250,000 gates, and up to 54 qubits but less than 500 gates. There are no circuits of large depth and width simultaneously. QASMBench \cite{qasmbench}, MQTBench \cite{mqtbench}, Feynman \cite{feynman_suite} and Veri-Q \cite{veri_q} also house limited numbers of wide, deep circuits -  of 2,901 circuits across the four suites, only 1 circuit is over 100 qubits wide and 1,000,000 gates deep. When the future of quantum computing is millions of qubits and gates, it is questionable whether these provide adequate scale for effective testing - whilst ideally circuits would be both deep and wide, there are only a few available which are either relatively deep, relatively wide, but not both.

\begin{figure}[tbp]
    \centering
    \includegraphics[width=1.0\linewidth, trim={0cm 0cm 0cm 0cm},clip]{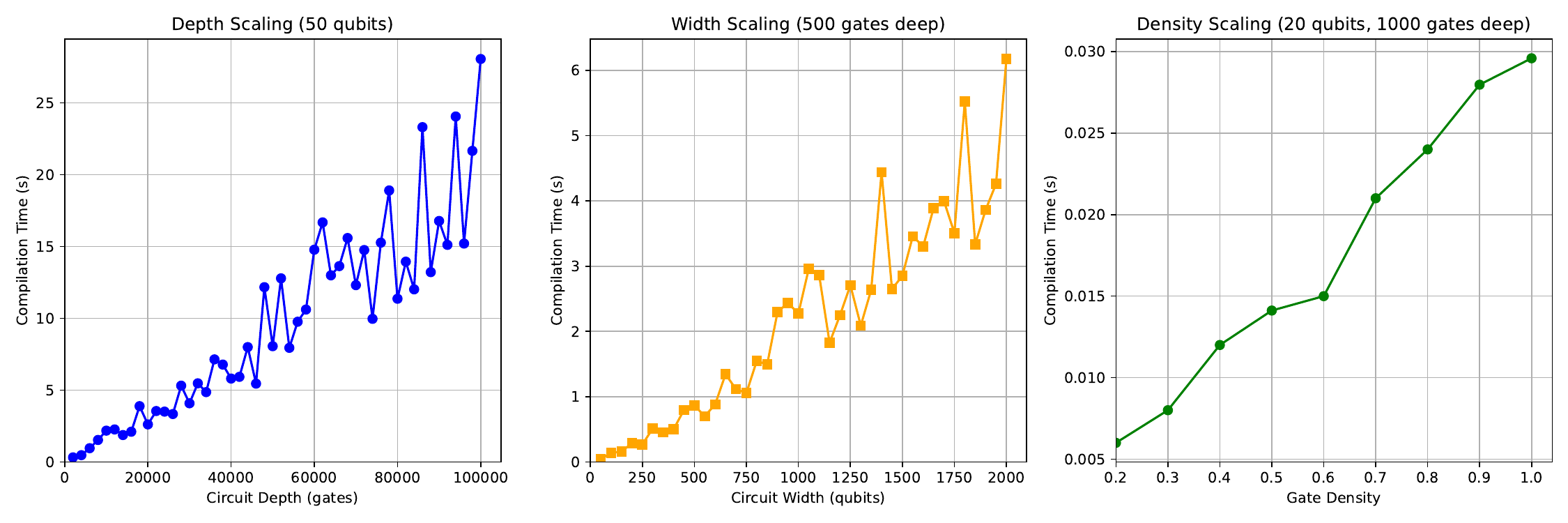}
    \caption{Compilation times for quantum circuits of varying depth, width and gate density.}
    \label{fig:compile-times-all}
\end{figure}

In light of this limitation on test circuits, random circuit generators - which allow generation of circuits of any combination of width and depth - may prove invaluable, yet these too suffer limitations. Specifically, QASMBench introduces the concept of gate density, representing the `occupancy' of a circuit i.e, the likelihood that a qubit is idle at a given time due to gate dependencies \cite{qasmbench}. The gate density is proportional to the number of gates in a circuit for a fixed depth and width. Gate density is a critical factor influencing compilation time, as shown in Fig. \ref{fig:compile-times-all}, which describes the growth in Qiskit's compilation time with circuit depth, width and density. 

A number of random circuit generators are available. The Qiskit generator allows control of the width and depth, but all the results have density of 100\%. The same is true of \cite{11009047}. This is problematic because it is not representative of hand-crafted circuits nor the densities of circuits in benchmark libraries. The gate densities of the five benchmark suites are illustrated in Fig. \ref{fig:combined_gate_depth_denisty_stats}b). The existing benchmark circuits are of lower density values, typically less than 60\%. If random circuit generation is to be used in lieu of a plentiful supply of benchmark circuits, then it is desirable that the circuits generated have similar characteristics to benchmarks, with respect to the factors influencing compilation time. This is not currently the case; current circuit generators do not allow control of density and hence this cannot be changed to match the densities of existing benchmark suites. 

This paper resolves these two issues. The Results section presents the experimental results for a proposed parallel compilation method, tested using the large-scale randomly-generated circuits alongside existing benchmark circuits. The Discussion section offers a further interpretation of the results. 
Finally, the Methods section also presents a method for generating random circuits with a prescribed gate density to allow generation of large-scale test circuits for parallel circuit compilation. This is used to generate large volumes of test circuits for the methodology later presented to allow parallelisation of the compilation process across multicore computing architectures. 
The parallel compilation is achieved through the decomposition of the original circuit, followed by it's compilation on independent processors. A permutation circuit is inserted to reset the qubit mappings and the sub-circuits are concatenated together to produce the final compiled circuit.

\section{Results}\label{sec:experiments}
The parallel compilation method is implemented in Python 3.10, using Qiskit 1.1.0 and PyTKET 2.3.2. The processor topology is a $2 \times m$ grid layout where $m=\left\lceil \frac{width}{2} \right\rceil$, chosen to be scalable and reflective of existing hardware, such as IBM Melbourne \cite{ibm_melbourne_device}. Experiments are executed on up to 16 cores of an AMD-based cluster, each a Dell PowerEdge R6525 server with dual AMD EPYC 7702 64-core processors. In total, up to 480GB of RAM was utilized across the cores.  

The parallel compilation approach uses, but is independent of existing compilers. Hence we analyse how its ability to accelerate compilation varies with a number of factors: conventional compiler, routing algorithm, circuit width, depth and density. In all cases we consider both the acceleration enabled and the quality of result, as measured by number of SWAP gates inserted and the total depth of the resulting circuit. We consider these for both benchmarking circuit suites - MQTBench, QASMBench and Red Queen, chosen as they give the best coverage in terms of circuit depth, width and density - and circuits produced by our random circuit generator. In all cases the number of available processors was varied between 2 and 16 (increments of 2).

\subsection{Qiskit Results}
We consider two different routing approaches: \textit{SabreSwap} and \textit{BasicSwap}. In all cases the initial layout was set to \textit{trivial} and optimisation level 0. Coupling maps were generated using Qiskit's grid method, edges undirected.

\begin{figure}[tbp]
\centerline{\includegraphics[width=0.5\linewidth, trim={0 0.25cm 0 0.25cm},clip]{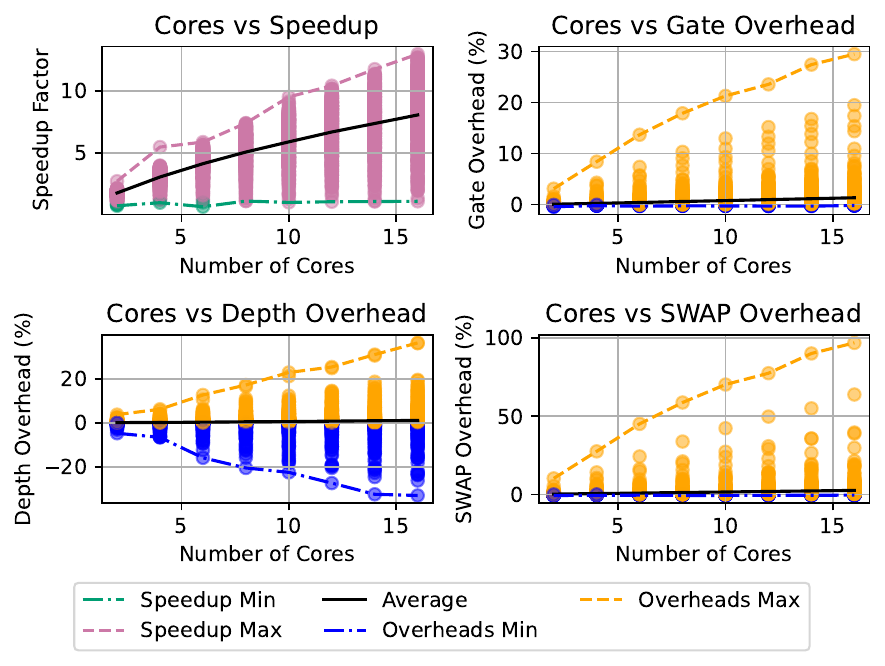}}
\caption{Speedup \& Overhead Variation with Number of Processors (Random Circuits, SabreSwap).}
\label{fig:regex_py_cores}
\end{figure}

For the both routing algorithm experiments a wide range of circuits were derived using our random circuit generator, ranging from 20 to 200 qubits (increments of 20) and depths of 10,000 to 100,000 gates (increments of 10,000). Each of these were derived with densities of 100\%, 70\%, 50\% and 20\%. In total, 400 circuits were generated. Fig. \ref{fig:regex_py_cores} shows how the speedup and overheads are influenced by the number of cores (and hence number of sub-circuits) for the SabreSwap algorithm. The dashed lines denote the absolute min and max values, and the solid line is the average.

Speedup is calculated from the wall time taken for the parallel and sequential compilation approaches. Time starts when the original QASM file is being read, and ends when the final compiled circuit is written to a QASM file.
Peak speedup for SabreSwap was 12.95, achieved when a circuit of 100\% density, 100 qubits and a depth of 100,000 is parallelised over 16 processors. The corresponding gate, SWAP and depth overheads were 0.03\%, 0.04\% and 0.41\%. Average speedup increases with the number of parallel units to 8.49 on 16 cores, with almost no relative increase in cost - overheads all remain less than 10\%. Maximum overhead values do grow with the number of cores, due to the impact of imposing permutation circuits on short circuits - this is entirely to be expected and indicates this approach is not suitable for short circuits (which, in the main, are not the key problem for parallelisation).

\begin{figure}[tbp]
\centerline{\includegraphics[width=1.0\linewidth, trim={0cm 0cm 0cm 0cm},clip]{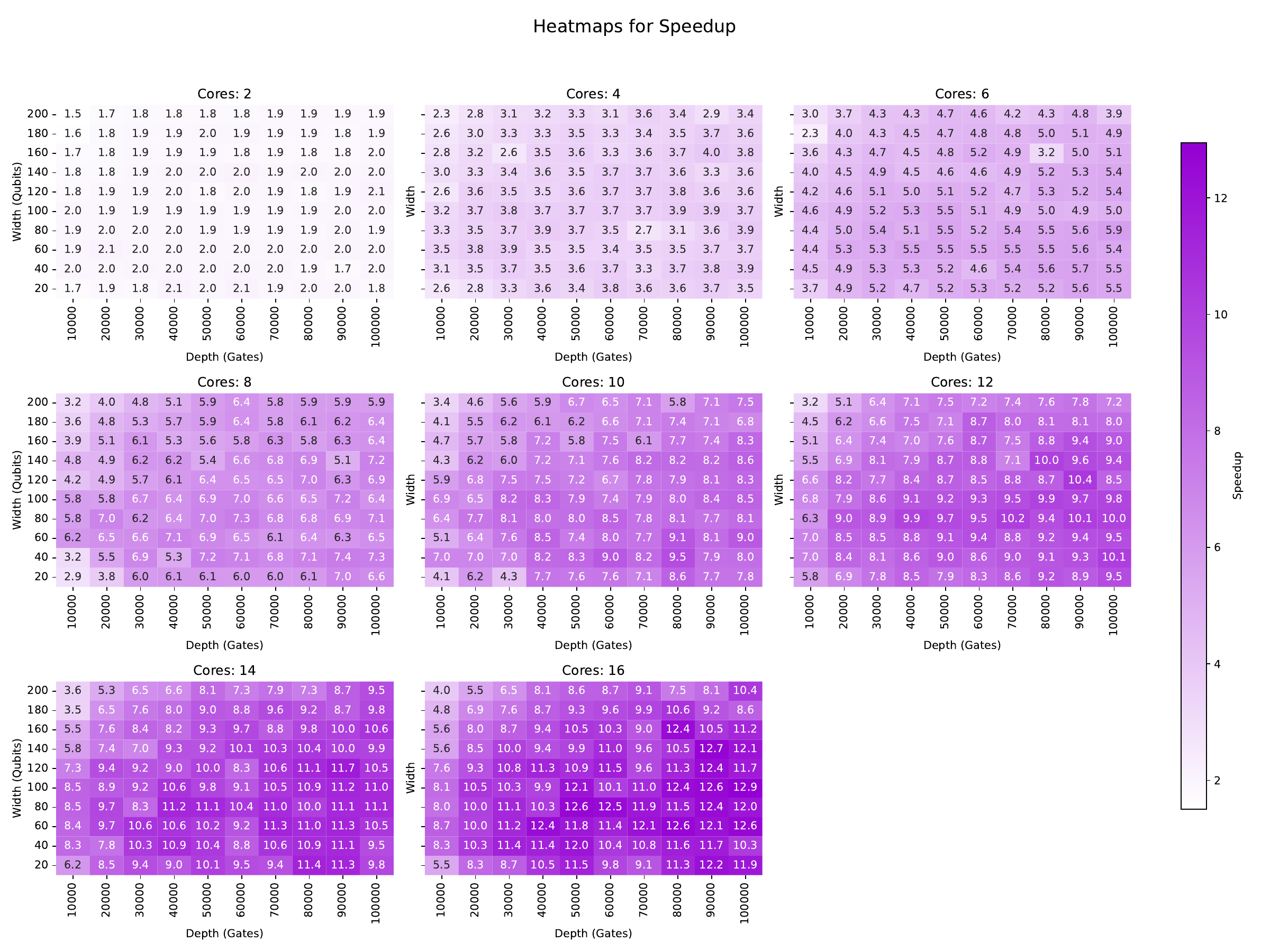}}
\caption{Speedup Heat Maps for 100\% Density Random Circuits using SabreSwap. }
\label{fig:regex_py_100_heatmap_Speedup}
\end{figure}

\begin{figure}[tbp]
\centerline{\includegraphics[width=1.0\linewidth, trim={0cm 0cm 0cm 0cm},clip]{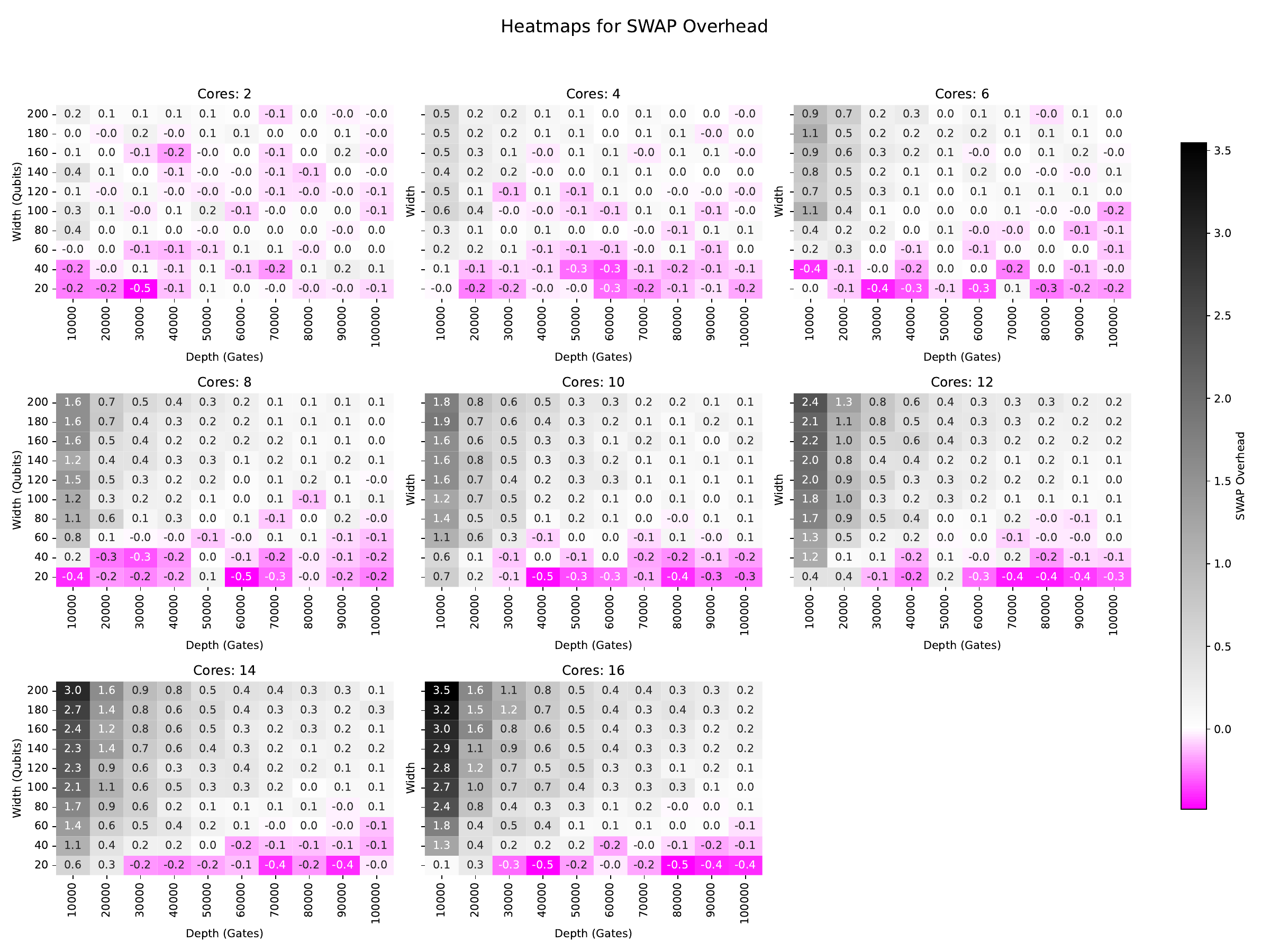}}
\caption{SWAP Overhead Heat Maps for 100\% Density Random Circuits using SabreSwap. }
\label{fig:regex_py_100_heatmap_SWAP}
\end{figure}

Fig. \ref{fig:regex_py_100_heatmap_Speedup} shows heatmaps for the speedup achieved using the parallel method over the sequential SabreSwap implementation. White indicates a minimal speed (at 1.5x faster) and the darkest purple represents the peak speedup. It can be seen that not only does the speedup increase with the number of cores, but it also increases with the circuit depth, i.e. deeper circuits see greater benefit from the sub-circuit approach. The also appears to be a correlation of the speedup achieved with the circuit width, up to a threshold for each configuration. This trend is not as well defined as the depth correlation however as the primary result of the sub-circuit decomposition is a reduction in the depth of the circuits being compiled. The width remains unaffected.

Fig. \ref{fig:regex_py_100_heatmap_SWAP} contains SWAP Overhead heatmaps for all 8 parallel configurations. Pink denotes reductions in SWAP cost, white no change and black the extreme overhead at an additional 3.5\% SWAP gates. As the number of sub-circuits increases, so does the number of permutation circuits and hence SWAPs. Overhead grows with number of processors and qubits but its impact is reduced for deeper circuits, indicating parallelisation should scale with both circuit depth and width - the ideal effect for the large scale circuits this paper considers.

\begin{figure}[tbp]
  \centering
  \begin{minipage}[b]{0.49\linewidth}
    \vspace{-3mm}
    \centerline{\includegraphics[width=1.0\linewidth]{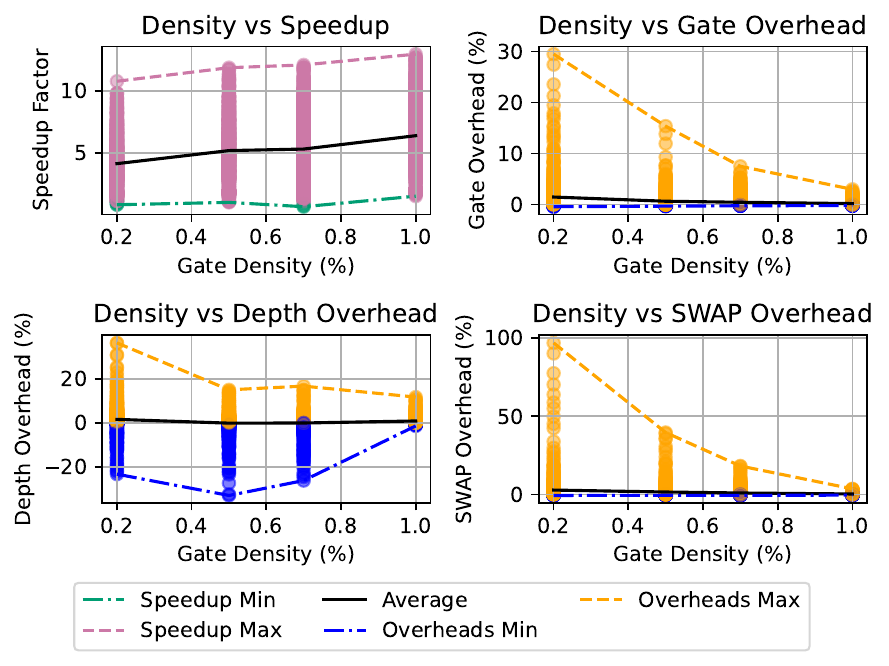}}
    \vspace{-5mm}
    \caption{Speedup \& Cost Variation with Random Circuit Density (SabreSwap).}
    \label{fig:regex_py_density}
  \end{minipage}
  \hfill
  \begin{minipage}[b]{0.49\linewidth}
    \vspace{-3mm}
    \centerline{\includegraphics[width=1.0\linewidth]{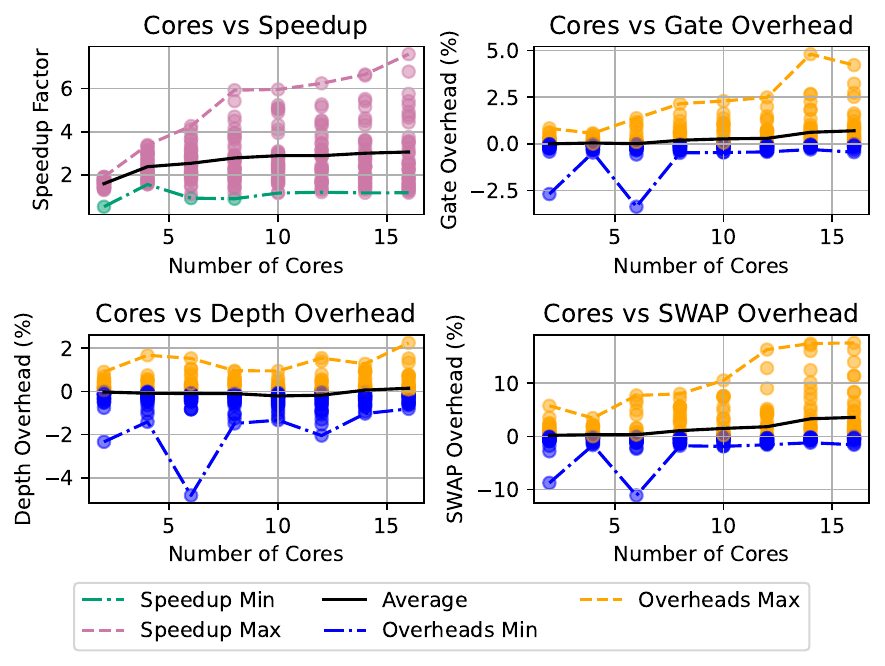}}
    \vspace{-5mm}
    \caption{Speedup \& Cost Variation with Number of Cores (Benchmark Circuits, SabreSwap).}
    \label{fig:regex_py_cores_bench}
  \end{minipage}
\end{figure}

Fig. \ref{fig:regex_py_density} shows how circuit density impacts speedup and overhead with SabreSwap. Speedup increases and overheads converge towards zero as gate density increases. Both may be linked to the increased number of gates. As the number of gates grows, so do routing times (in a super-linear fashion), hence parallelising into multiple, smaller circuits leads to significant speedups. Further, as gate density increases, the gate count grows and the relative impact of the additional permutation circuits reduces.

\begin{figure}[tbp]
  \centering
  \begin{minipage}[b]{0.49\linewidth}
    \centerline{\includegraphics[width=1.0\linewidth]{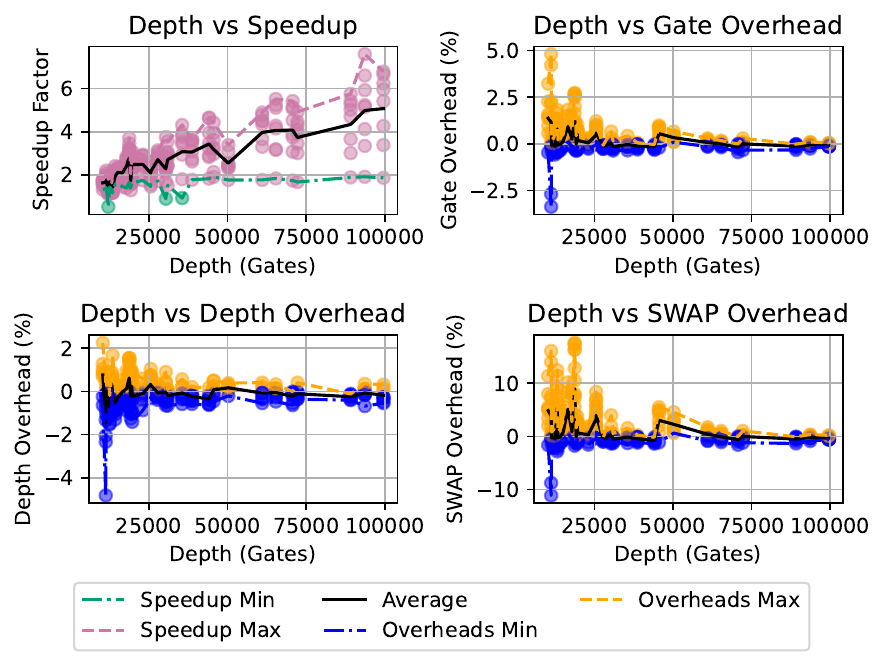}}
    \vspace{-5mm}
    \caption{Speedup \& Cost Variation with Benchmark Circuit Depth (SabreSwap).}
    \label{fig:regex_py_depth_bench}
  \end{minipage}
  \hfill
  \begin{minipage}[b]{0.49\linewidth}
    \vspace{-3mm}
    \centerline{\includegraphics[width=1.0\linewidth]{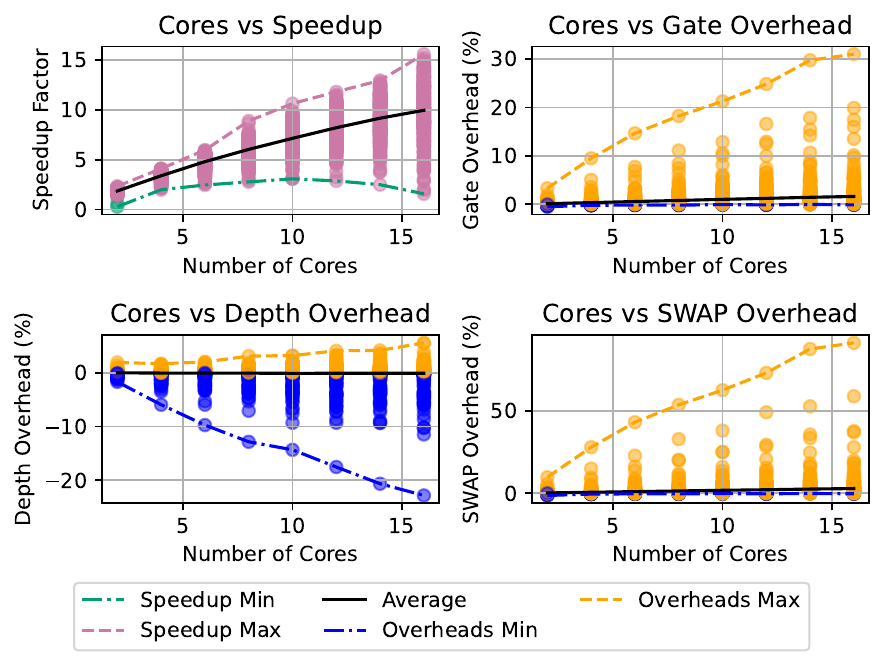}}
    \vspace{-5mm}
    \caption{Speedup \& Cost Variation with Number of Processors (random circuits, BasicSwap).}
    \label{fig:basic_cores}
  \end{minipage}
\end{figure}

The 36 benchmark circuits have depths of 10,000 - 100,000 gates widths of 8 - 29 qubits. Their densities vary between 4.43\% - 33.12\%. Fig. \ref{fig:regex_py_cores_bench} shows that speedup increases up to 8 cores; this is due to the small number of qubits in the test circuits and the limited gate density limiting the scope. Average gate, SWAP and depth overheads all remain near zero. Peak speedup was 7.58, achieved on 16 sub-circuits for a circuit with 15 qubits and a depth of 93,645 gates. Fig. \ref{fig:regex_py_depth_bench} shows that trends in speedup and overhead variation with depth match those of the random circuits, accounting for the more limited scale of the benchmark circuits. The limited variation in width or density of these benchmarks precludes useful analysis of any resulting trends.

The same 400 random circuits were compiled using the parallel BasicSwap implementation. 
The results with the BasicSwap algorithm are shown in Fig. \ref{fig:basic_cores}, which shows this reflects the SabreSwap trends. Peak speedup of 15.56 was achieved for 16 sub-circuits, 100 qubits at a depth of 30,000 gates and density of 100\%. 

\begin{figure}[tbp]
  \centering
  \begin{minipage}[b]{0.49\linewidth}
    \vspace{-3mm}
    \centerline{\includegraphics[width=1.0\linewidth]{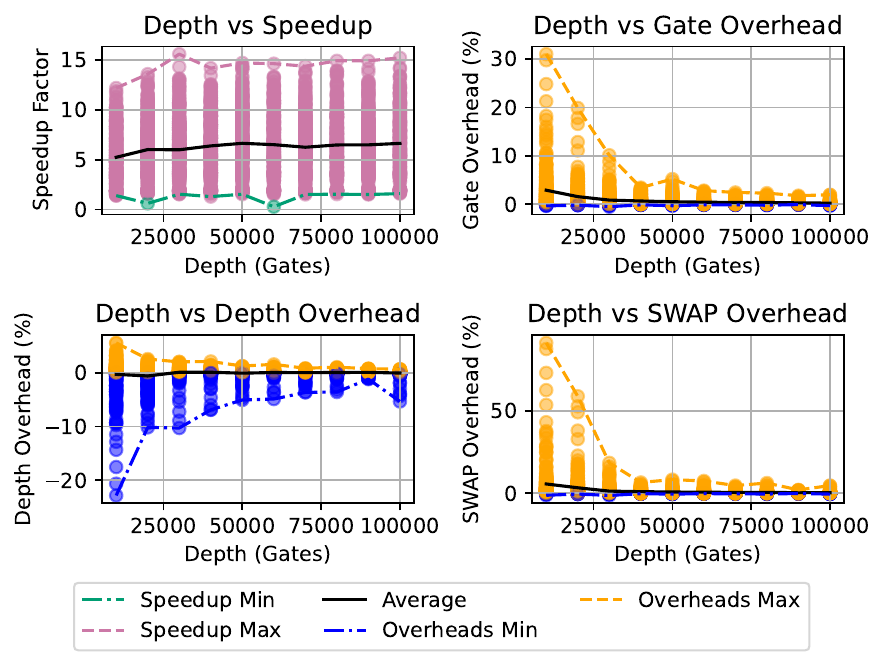}}
    \vspace{-5mm}
    \caption{Speedup \& Cost Varying with Random Circuit Depth (BasicSwap).}
    \label{fig:basic_depth}
  \end{minipage}
  \hfill
  \begin{minipage}[b]{0.49\linewidth}
    \centerline{\includegraphics[width=1.0\linewidth]{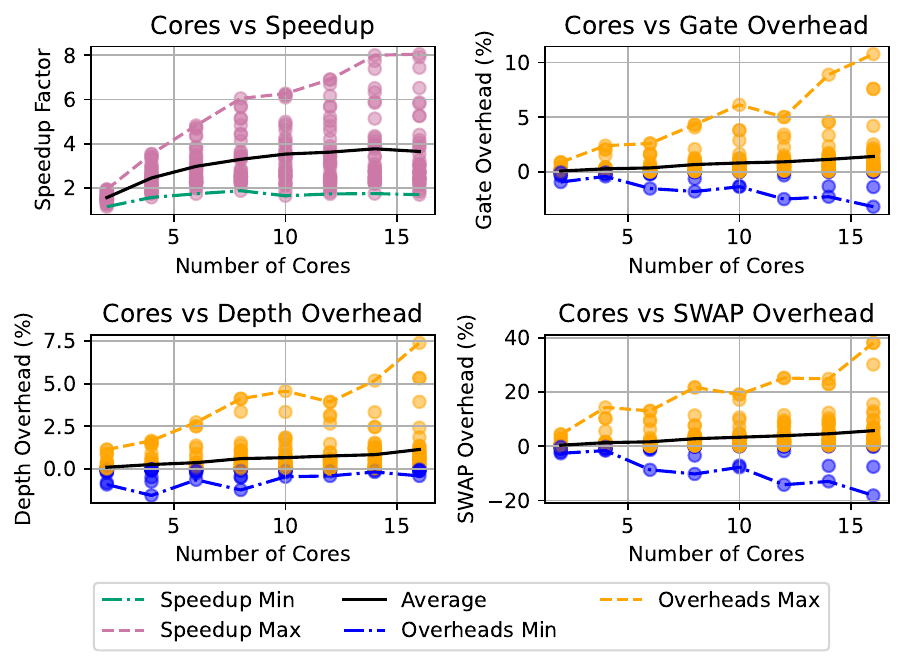}}
    \vspace{-5mm}
    \caption{Speedup \& Cost Variation with Number of Processors (benchmark circuits, BasicSwap).}
    \label{fig:basic_cores_bench}
  \end{minipage}
\end{figure}

Fig. \ref{fig:basic_depth} shows speedup for BasicSwap remains approximately uniform with increasing depth; initially cost varies but as depth increases overheads all tend towards 0\%. The initial variation is less extreme than that observed for SabreSwap, and the convergence more rapid. Alongside higher initial speedup, this suggests BasicSwap sees the benefit of parallelisation more readily than SabreSwap. 
The influence of the width and density on the speedup and overheads reflect the trends seen with SabreSwap.

\begin{figure}[tbp]
  \centering
  \begin{minipage}[b]{0.49\linewidth}
      \centerline{\includegraphics[width=1.0\linewidth]{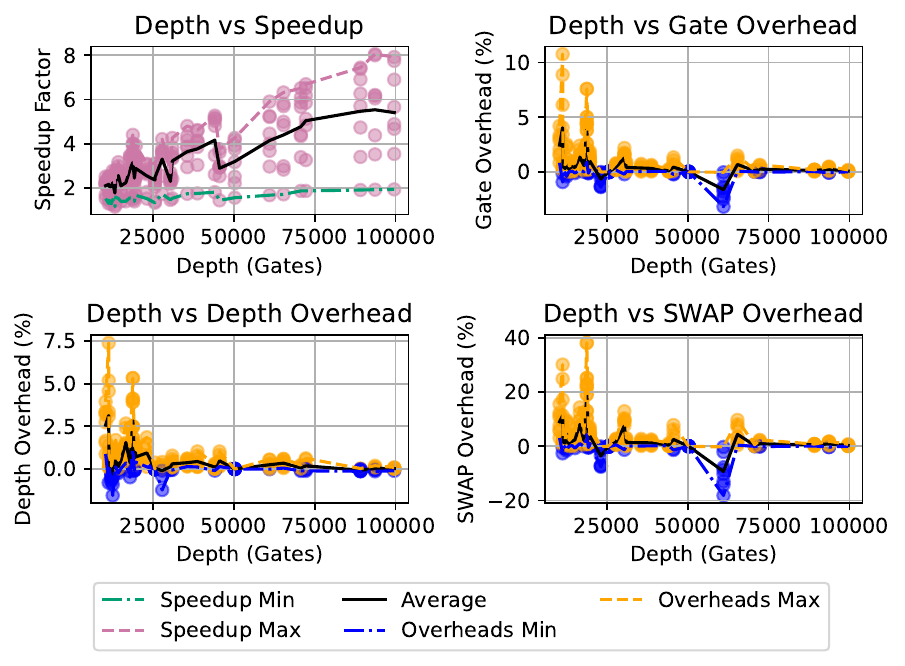}}
    \vspace{-5mm}
    \caption{Speedup \& Cost Varying with Benchmark Circuit Depth (BasicSwap).}
    \label{fig:basic_depth_bench}
  \end{minipage}
  \hfill
  \begin{minipage}[b]{0.49\linewidth}
      \vspace{-3mm}
    \centerline{\includegraphics[width=1.0\linewidth]{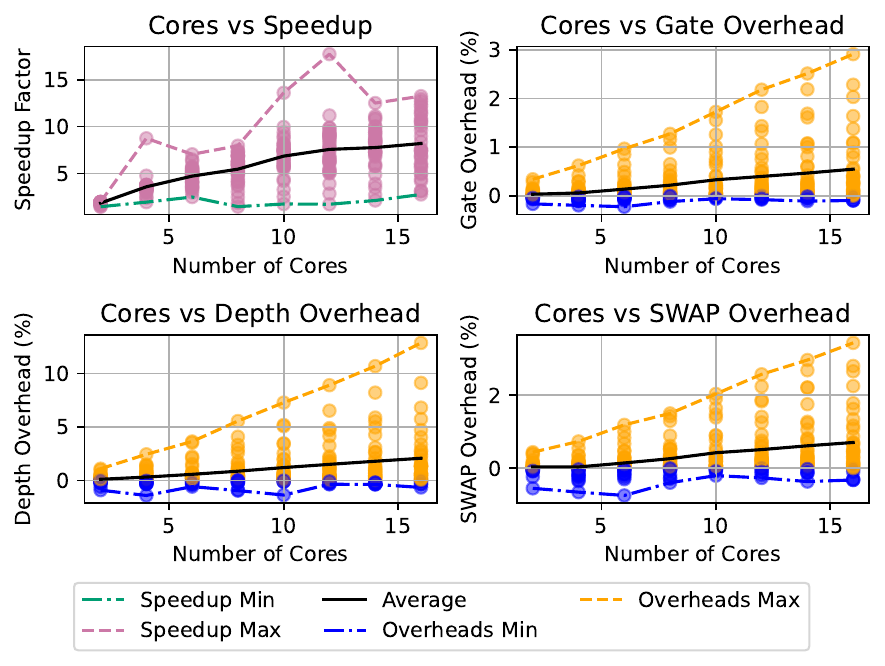}}
    \vspace{-5mm}
    \caption{Speedup \& Cost Variation with Number of Processors for Qiskit 2.2.3 (SabreSwap).}
    \label{fig:qiskit2_cores}
  \end{minipage}
\end{figure}

When applied to the same 36 benchmark circuits, Fig. \ref{fig:basic_cores_bench} shows that varying the number of parallel units increases speedup up to 12 processors. Peak speedup of 8.05 was achieved using 16 sub-circuits for urf6\_160 \cite{red-queen}. Fig. \ref{fig:basic_depth_bench} shows that varying depth results in the same trends observed for the randomly-generated circuits - the average speedup increases and overheads tend towards 0\%. The extreme values for the SWAP count overhead were greater than the extremes seen with SabreSwap due to the less optimal nature of the BasicSwap SWAP insertion algorithm.

\subsection{Performance with Qiskit 2.2.3}
The large scale experiments for both algorithms have been run using Qiskit 1.1.0 as which was the stable release at the onset of the experiments. To account for updates in the Qiskit framework, a subset of experiments using Qiskit 2.2.3, the most recent release at the time of writing, have been executed. These experiments use SabreSwap and have only be run for the 100\% density randomly generated circuits due to time constraints. The results are shown in Fig. \ref{fig:qiskit2_cores}. % and Fig. \ref{fig:qiskit2_depth}. 
As can be seen, the trends observed using Qiskit version 1.1.0 are consistently reproduced in the results obtained with version 2.2.3, irrespective of upgrades to the Qiskit Transpiler. Improvements in transpiler performance, in particular with respect to compilation time, are directly reflected in the parallel approaches ability to capitalize on these improvements. The main impact of such improvements is a potential shift in the effective speedup thresholds for circuit depth, width, and density, consistent with effects observed with other compilers later in this section.

\subsection{PyTKET Results}
The same experiments were repeated for PyTKET 2.3.2 and RoutingPass \cite{Pytket_qubit_routing}. To apply the trivial initial layout, a placement pass is used to locate qubits according to the mapping given. Lookahead depth was the default value of 10 (\url{https://docs.quantinuum.com/tket/api-docs}).

Fewer random circuits were explored using PyTKET, owing to constraints and the longer runtime of its compiler compared to Qiskit (time limit for each circuit was set to 3 days). These circuits ranged in width from 20 - 100 qubits (increments of 20), 10,000 - 100,000 gates (increments of 10,000) and densities of 100\%, 70\%, 50\% and 20\%.

\begin{figure}[tbp]
  \centering
  \begin{minipage}[b]{0.49\linewidth}
    \centerline{\includegraphics[width=1.0\linewidth]{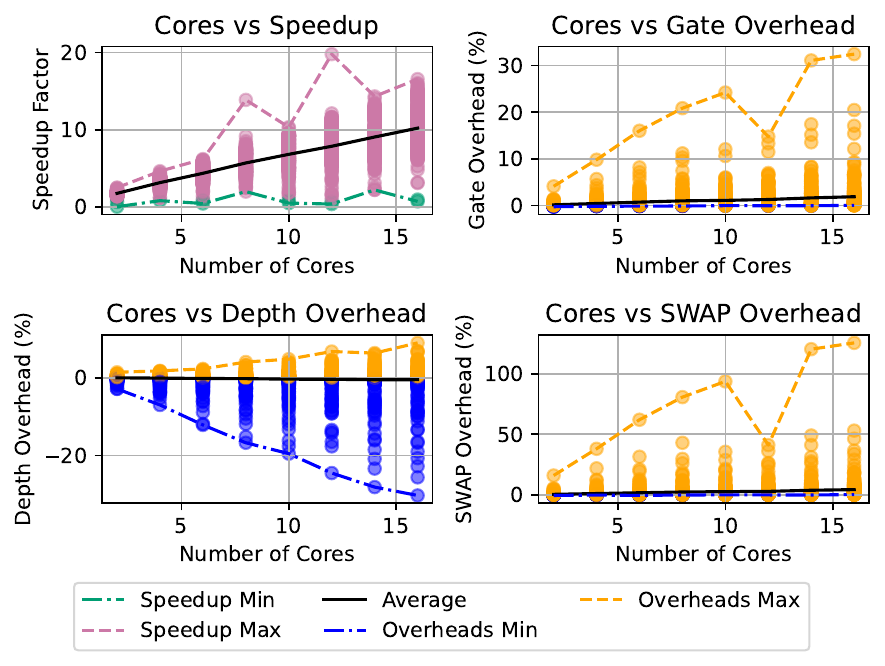}}
    \vspace{-5mm}
    \caption{Speedup \& Cost Variation with Processors (random circuits, PyTKET).}
    \label{fig:Pytket_cores}
    \end{minipage}
    \hfill
    \begin{minipage}[b]{0.49\linewidth}
    \centerline{\includegraphics[width=1.0\linewidth]{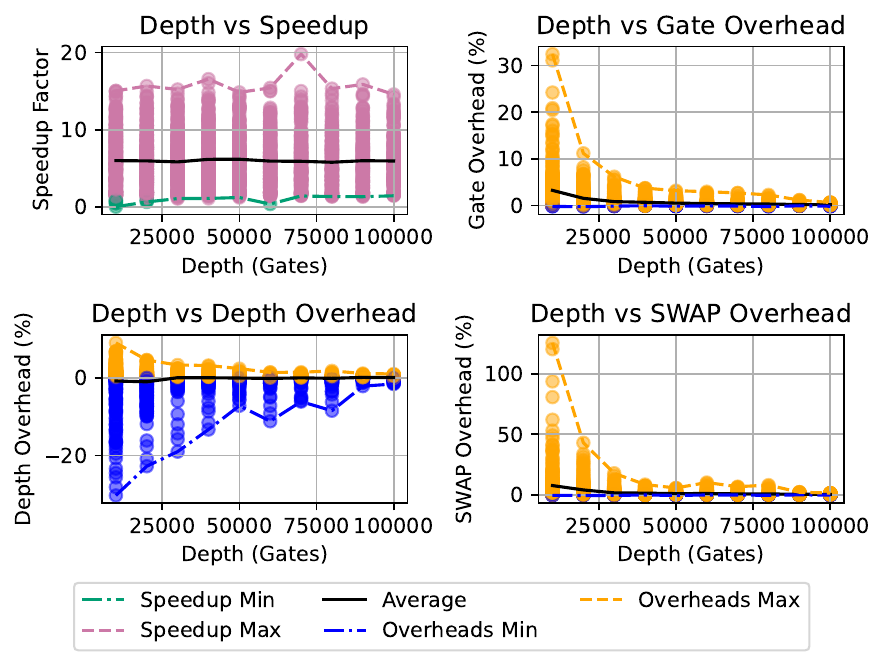}}
    \vspace{-5mm}
    \caption{Speedup \& Cost Variation with Random Circuit Depth (PyTKET).}
    \label{fig:Pytket_depth}
  \end{minipage}
\end{figure}

Fig. \ref{fig:Pytket_cores} shows that speedup average speedup increases with the number of sub-circuits and that average gate, SWAP and depth overheads are negligible. Peak speedup achieved was 19.80 with a corresponding gate, depth and SWAP overheads of 0.17\%, 0.23\% and 0.25\%. Fig. \ref{fig:Pytket_depth} shows that speedup appears to remain constant for increasing circuit depth, suggesting that the length of circuit required to observe any speedup is less than 10,000 gates. Hence, each different compilation algorithm has a different optimum number of sub-circuits for circuits of given physical attributes.

\begin{figure}[tbp]
  \centering
  \begin{minipage}[b]{0.49\linewidth}
    \centerline{\includegraphics[width=1.0\linewidth]{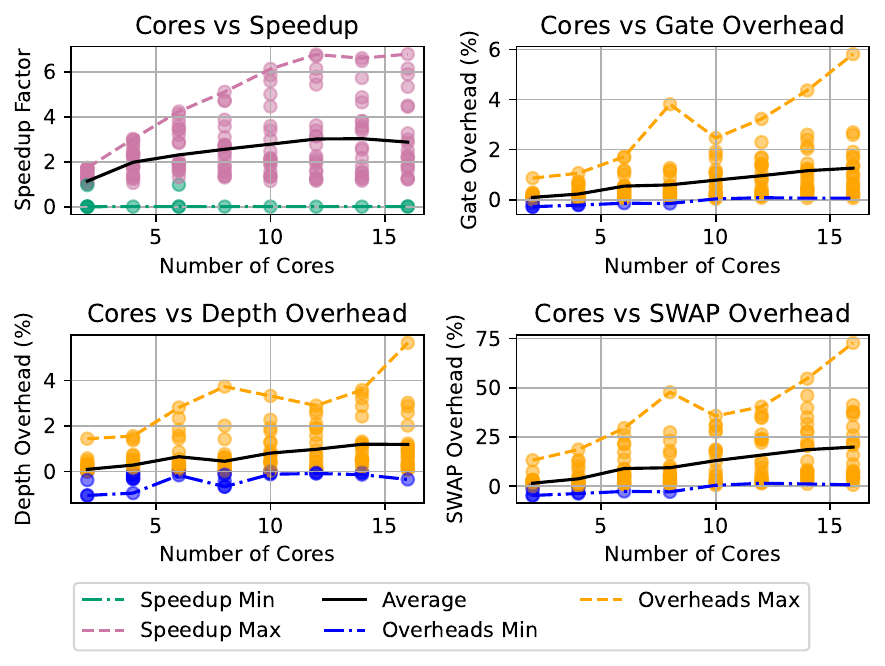}}
    \vspace{-5mm}
    \caption{Benchmark Speedup \& Cost Variation with Processors (PyTKET).}
    \label{fig:Pytket_cores_bench}
    \end{minipage}
    \hfill
    \begin{minipage}[b]{0.49\linewidth}
    \vspace{-3mm}
    \centerline{\includegraphics[width=1.0\linewidth]{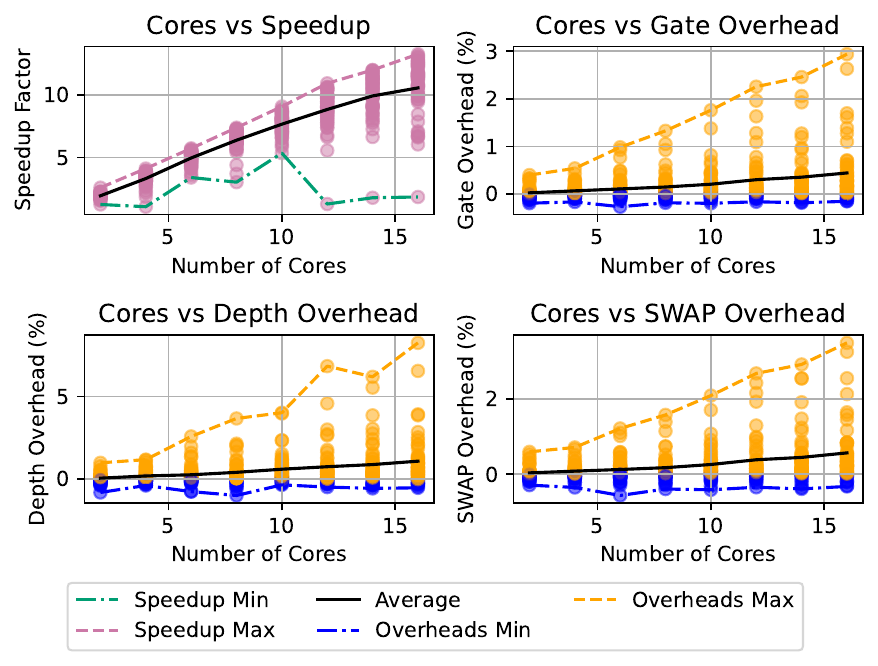}}
    \vspace{-5mm}
    \caption{Speedup \& Cost Variation with Number of Processors for a linear processor (Qiskit, SabreSwap).}
    \label{fig:linear_cores}
  \end{minipage}
\end{figure}

When used to compile the benchmark circuits, the results can be seen in Fig. \ref{fig:Pytket_cores_bench}. The peak speedup achieved was 6.79 with 16 cores on the same benchmark circuit that achieved the peak speedup for both Qiskit algorithms (urf6\_160). Again the results for PyTKET behave similarly to the results for the Qiskit algorithms. The use of a different compiler demonstrates the versatility of the parallel approach proposed.

\subsection{Performance With a Linear Processor Shape}
The grid-shaped processor coupling-map used in the large-scale experiments enjoys a high level of qubit connectivity with each qubit having up to 4 qubits connected directly to it. Current trends in processor development appear to focus on sparser connections i.e. a lower qubit connectivity, such as with heavy-hex \cite{topological_and_subsystem_codes} \cite{who_is_leading}. Taking this to the extreme, a subset of the of SabreSwap experiments have been repeated using a 1-D linear processor shape (i.e. each qubit is connected to at maximum 2 other qubits), with Qiskit's SabreSwap algorithm for the 100\% density circuits for up to 100 qubits. The results are shown in Fig. \ref{fig:linear_cores}. % and Fig. \ref{fig:linear_depth}. 
The results reproduce the trends observed with the grid processor, suggesting that that parallel approach is effective for a range of processors with different levels of qubit connectivity.

\subsection{Memory Usage Analysis}
\begin{figure}[tbp]
\centerline{\includegraphics[width=0.9\linewidth]{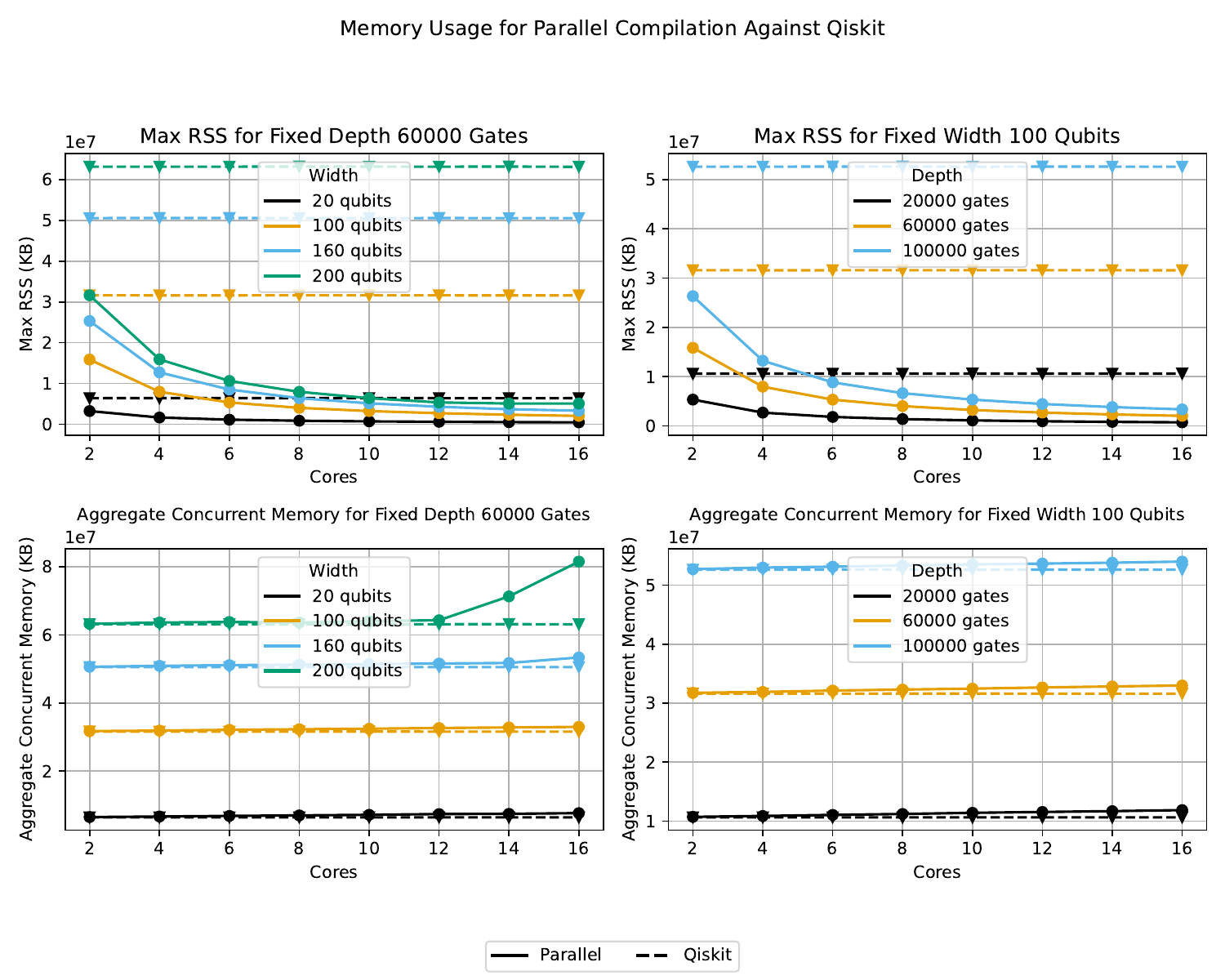}}
\caption{Memory usage analysis using Qiskit's SabreSwap algorithm.}
\label{fig:memory_analysis}
\end{figure}

In addition to consideration of the quantum overheads incurred by the parallel compilation process, the computational memory should also be considered. This was reported using the GNU \verb|/usr/bin/time -v| utility. Each phase of execution (decomposition, compilation and concatenation) was wrapped in the command to record the maximum resident set size (MaxRSS) reached during that phase.

For the parallel implementation, the reported MaxRSS corresponds to the peak per‑process memory usage among all concurrently executing compilation tasks. To estimate the aggregate concurrent memory footprint, relevant for HPC node schedulability under Slurm, we multiply the MaxRSS by the number of cores. Qiskit’s memory usage for the non‑parallel (monolithic) baseline was measured using the same procedure, and all runs were executed on the same HPC cluster under identical conditions.

Fig. \ref{fig:memory_analysis} shows the peak memory consumption for multiple compilation runs from 2 to 16 cores. As can be seen the memory requirements increase with increasing circuit size as would be expected. However the top 2 plots show the peak memory required for the parallel implementation starts lower and as the number of cores, and hence sub-circuits is increased, the peak memory required decreases.

The aggregated memory requirements for the parallel and monolithic approaches remain similar for increasing depth and cores as shown on the bottom right plot. Similarly, for circuits of up to 100 qubits this is also the trend for increasing circuit width. For circuits of width greater than 100 qubits, with larger numbers of cores, there begins to appear a greater memory requirement for the parallel approach. This indicates that for wide circuits, care must be taken when selecting the number of cores: the speedup gained must be balanced against the increased memory overhead introduced by parallelisation, as well as the quantum gate and depth overheads explored previously.

\subsection{Circuit Behaviour Analysis}
Although our circuit decomposition approach does not cut gates and preserves the global gate ordering by strictly concatenating compiled sub-circuits in the original sequence, the parallel compilation strategy may restrict the opportunity for cross-sub-circuit optimisations such as gate commutation or cancellation \cite{gate_commutation}. These limitations can affect circuit depth, gate count, or local gate patterns, even when functional correctness is maintained.

In addition to analysing compilation speedup and overhead, it is important to evaluate if the behaviour of the original quantum circuit is preserved in the parallel compilation process. To assess this we performed an additional study to quantify the fidelity of the circuit produced through the parallel compilation in comparison to the circuit produce by the monolithic compilation. Benchmark circuits were compiled using the parallel compilation method, with Qiskit's SabreSwap routing algorithm and the same circuits were compiled monolithically using SabreSwap for baseline comparison. The resulting compiled circuits were simulated on a noiseless backend together with the original, non-compiled circuit. 

\begin{figure}[tbp]
\centerline{\includegraphics[width=\linewidth]{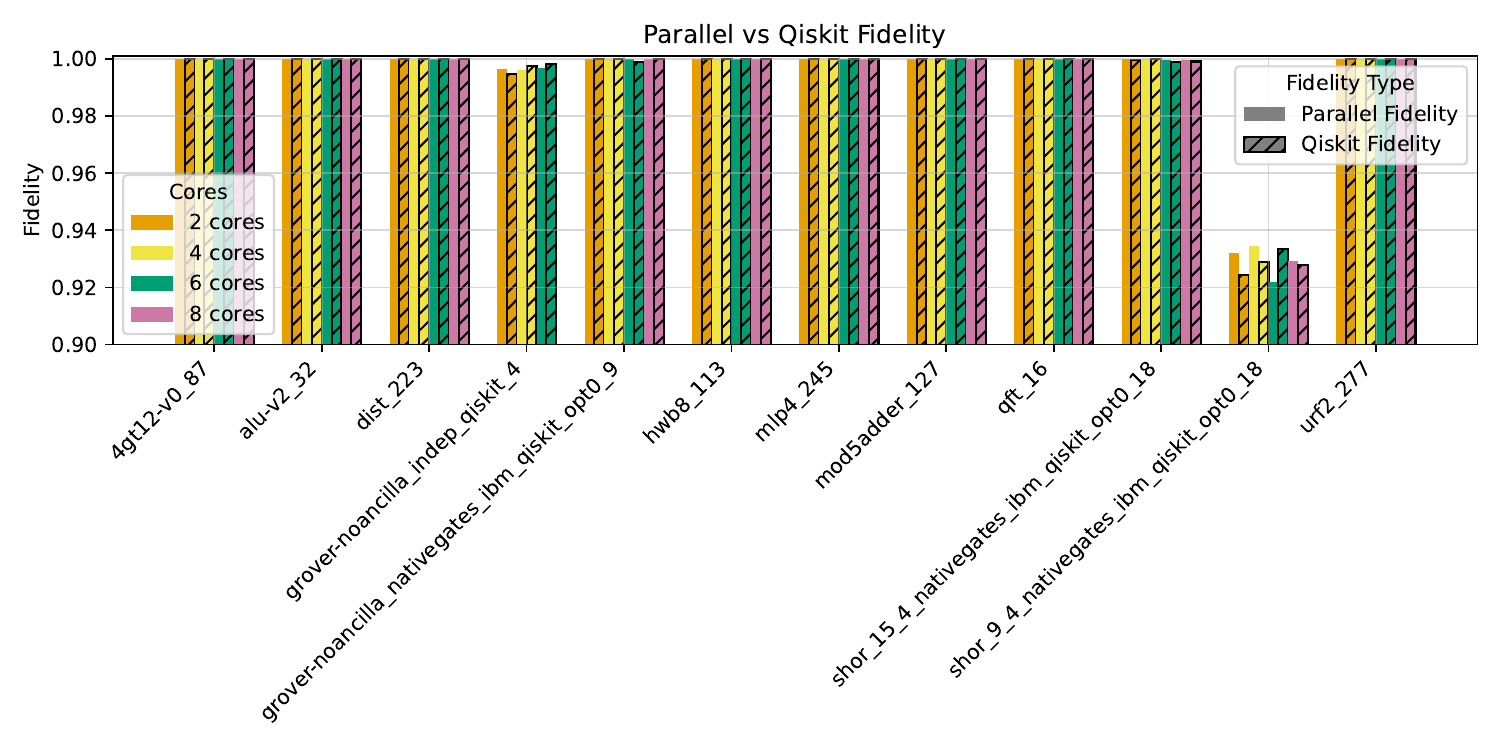}}
\caption{Fidelity analysis using Qiskit's SabreSwap algorithm both monolithically and with the parallel implementation.}
\label{fig:fidelity_analysis}
\end{figure}

For each benchmark we calculated (a) the fidelity between the original circuit and the parallel-compiled output, and (b) the fidelity between the original circuit and the monolithic Qiskit output. The fidelity results are shown in Fig. \ref{fig:fidelity_analysis}. As can be seen, the parallel‑compiled circuits achieve fidelity values closely matching those of the monolithic compilation. This indicates that the parallel compilation approach does not introduce meaningful behavioural deviations and preserves the intended circuit transformation while enabling substantial compilation speed improvements.

\subsection{Random Circuit Generator Analysis}
The final open question resulting from the work in this paper concerns the effectiveness of the random circuit generation technique for benchmarking compiler performance and cost, relative to that which would be expected from using a recognised benchmarking suite. For instance, when comparing Fig. \ref{fig:regex_py_cores} (profiling circuits produced by the random circuit generator) and Fig. \ref{fig:regex_py_cores_bench} (circuits from the benchmark suites), there are noticeable differences in average and maximum speedup in particular, as well as extreme values for all 3 overhead measures. An open question, however, is whether this is due to the different scales of circuits produced by the generator, or whether there is a mismatch in some aspect of circuit structure which causes this discrepancy. The maximum width of the random circuits is 200 qubits, whereas the maximum depth in the benchmark circuits is 29 qubits, which is much smaller. In addition the average gate density of the benchmark circuits ranges from 4.43\% to 33.12\%, compared to the random circuits which range from 20\% to 100\%.

\begin{figure}[tbp]
  \centering
  \begin{minipage}[b]{0.49\linewidth}
    \vspace{-3mm}
    \centerline{\includegraphics[width=1.0\linewidth]{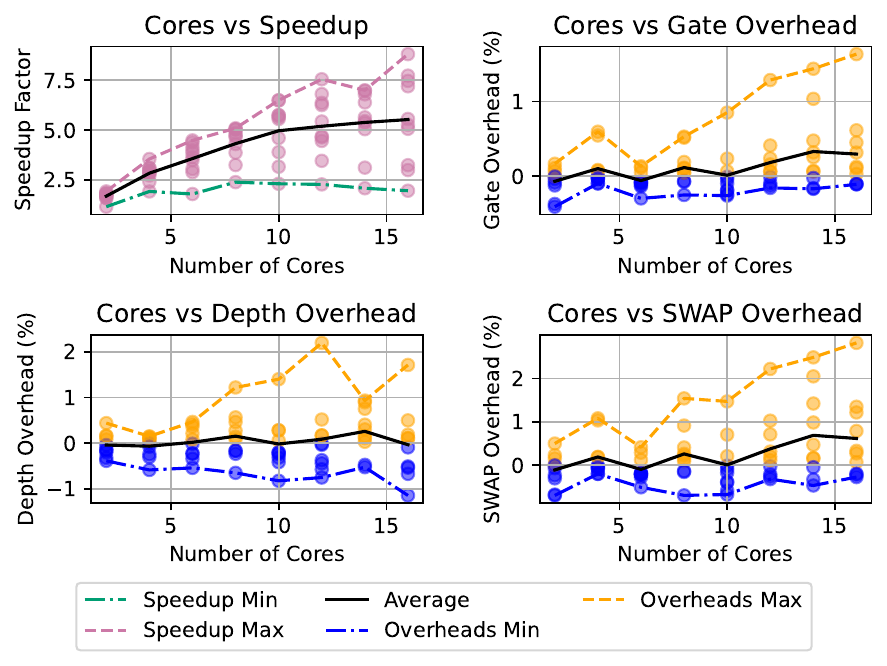}}
    \vspace{-5mm}
    \caption{Recorded speedup and overheads for the 20-qubit randomly-generated circuits at 20\% density.}
    \label{fig:regex_py_cores_20_dens_20q}
  \end{minipage}
  \hfill
  \begin{minipage}[b]{0.49\linewidth}
    \centerline{\includegraphics[width=1.0\linewidth]{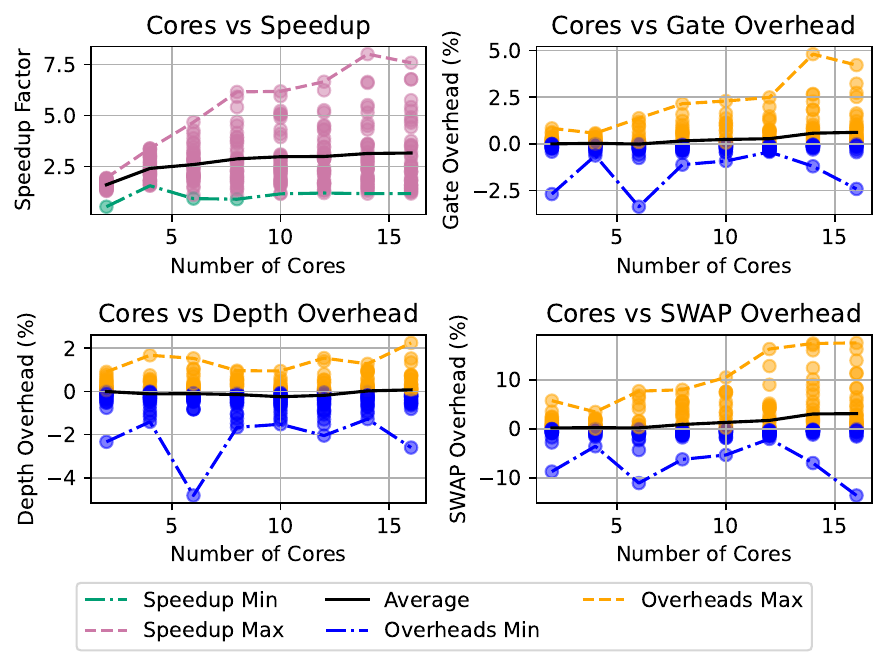}}
    \vspace{-5mm}
    \caption{Speedup \& Cost Variation with Number of Processors for benchmark circuits, including VQE algorithm.}
    \label{fig:vqe_cores}
  \end{minipage}
\end{figure}

In an attempt to shed light on this question, we compare random circuits generated to meet the same width, height and density characteristics as the benchmarks. The average width of the benchmark circuits is 14 qubits and the average gate density is 19.80\%. The previous random circuit results have been filtered to show the results for 20-qubit circuits of 20\% density. These filtered results are shown in Fig. \ref{fig:regex_py_cores_20_dens_20q}. When compared to the benchmark results in Fig. \ref{fig:vqe_cores} - which shows the previous SabreSwap results, now supplemented with the VQE algorithm for comparison with the random circuit generator - the scale of speedup and overheads are more closely replicated. 
The benchmark circuits contain results for circuits widths as low as 8 qubits and densities as low as 4.43\%, hence there is still some disparity between the results, however the 20\% circuits are a better representation of the benchmark circuits than the higher density circuits such as the 100\% density circuit produced from the Qiskit random circuit generator.

The benchmark suites include circuits derived from well‑known quantum algorithms such as Shor’s algorithm and Grover’s algorithm \cite{shor} \cite{grover}. Although the current approach to generating low‑density random circuits does not explicitly reproduce the structured characteristics of such algorithms, extending these methods to better capture algorithm‑specific structure represents a promising direction for future work.

\section{Discussion}\label{sec:discussion}
Quantum compilation is a bottleneck in the quantum computing pipeline, with compile times for future circuits, numbering very large numbers of qubits and very large numbers of gates, a major future problem. Further, even having access to benchmarks to test compilation techniques for very large circuits is currently a major limiting factor. This paper resolves this issue by proposing a method of generating random circuits which allow control of not only their width and depth but also their density, a third major factor influencing compilation time. 

We use this to generate over 8,000 circuit configurations for testing of fast compilation techniques on large circuits, alongside known benchmarks. We use these to test a new approach to parallelising the compilation problem by subdividing a circuit, compiling each sub-circuit in parallel, and reconstructing the original via permutation circuits to interface the sub-circuits. This approach is agnostic of circuit structure, compiler or routing algorithm.

When tested using Qiskit employing SabreSwap, peak speedup reached 12.95 and 6.38 on average. Average gate, SWAP and depth overheads were 0.2\%, 0.2\% and 0.85\% respectively. BasicSwap achieved a peak speedup of 15.56 with 16 cores with negligible overheads. When applied to TKET, speedup reached 19.80 and 9.09 on average. The use of three different routing algorithms across two different compilers shows the flexibility of the proposed approach while the large-scale experiments demonstrate the scalability of both the low density random benchmark suite, and the parallel compilation approach.

This level of testing comfortably exceeds that which has been attempted before. Specifically, it has extended the number of different types of circuits used to test a compiler from a previous high of 26, to 400. It has extended the previous maximum number of circuit instances from 500  to 8000. It has doubled the maximum number of qubits used (100 to 200) and increased the maximum number of gates from 100,000 to over 16 million.

The experimental results suggested that the speedup achieved was proportional to the number of sub-circuits, however the optimal number of sub-circuits is proportional to the physical circuit properties i.e. the depth, width and gate density. Future work may consider a model to determine the optimal number of sub-circuits for a circuit based on its physical attributes and compiler to be used. The use of a large number of randomly-generated circuits of varying depth, width and density may be used to achieve this. 

Additionally, the size of the sub-circuits is determined by the total number of gates in the circuit. The SWAP insertion component of compilation is often the most time consuming due to the NP-Complete complexity. SWAP insertion only considers multi-qubit gates and hence it may be beneficial to consider balancing the number of multi-qubit gates in circuits, rather than the total number of gates. This may lead to more optimal load balancing and greater speedup gains.

\section{Methods}
\subsection{Generating Random Circuits with a Given Gate Density}\label{sec:random_circ_gen}

The $depth$ of a circuit is the length of the critical path through the circuit. The $width$ of a circuit is the number of qubits in a circuit. The gate density, $d$, of a circuit of given depth and width can be defined as follows where $n_{q1}$ is the number of one-qubit gates, $n_{q2}$ is the number of two-qubit gates \cite{qasmbench}:
\begin{equation} \label{eq:density}
d = \frac{n_{q1} + (2 * n_{q2})}{depth * width}
\end{equation}

A circuit of given gate density can be generated by first using Qiskit's random circuit method to produce a circuit of 100\% density. Gates are iteratively removed from the circuit to achieve the target gate density. A random circuit of reduced gate density can be generated by using Algorithm \ref{alg:density} as follows:

\begin{itemize}
    \item The inputs: $depth$ is the circuit depth in gates, $density$ is the target gate density and $width$ is the number of qubits.
    \item Lines 2-4 generate a fully populated circuit of 100\% density using the Qiskit random circuit generator. The maximum number of operations is calculated and hence the number of operations required to achieve the desired density for the set depth and width is calculated. 
    \item Lines 5-7 choose a `safe qubit' at random. A random number is chosen for $n_{q1}$ and (\ref{eq:density}) is then used to calculate $n_{q2}$. 
    The safe qubit maintains the circuit depth as at the minimum density the safe qubit will have $depth$ 1-qubit operations. There will be no other operations in the circuit. Hence the minimum possible density is $\frac{1}{width}$.
    \item Lines 8-9 check the circuit to ensure that there are enough 1 and 2 qubit gates not operating on the safe qubit to facilitate the gate removal counts. If not, a different combination of removal counts is chosen, or the safe qubit is changed. Once verified, the safe qubit may be stripped from the original circuit to reduce the length of the instruction list.
    \item Lines 10-30 iterate through a while loop until the set quotas of 1 and 2-qubit gates are removed. For each iteration, a gate is chosen at random from the modified list of instructions, ensuring the safe qubit is not impacted. If the quota for the gate type is greater than 0, the gate is removed from both the original circuit and the modified instruction list. The gate quota is decremented.
    The corresponding number of qubit operations is appended to the removed operations count.
    The process repeats until both removal quotas reach zero.
\end{itemize}

\begin{algorithm}
\caption{Generate Low Density Random Circuits}
\begin{algorithmic}[1]\label{alg:density}
\STATE \textbf{Input:} $depth$, $width$, $density$
\STATE $random\_qc \gets$ random circuit of $depth$ gates deep and $width$ qubits
\STATE $max\_ops \gets depth*width$
\STATE $ops\_to\_remove \gets max\_ops - \lceil max\_ops * density \rceil$
\STATE $safe\_qubit \gets$ random int between $0$ and $width$
\STATE $n_{q1} \gets$ random int between $0$ and $max\_ops$
\STATE $n_{q2} \gets \frac{max\_ops - n_{q1}}{2}$
\STATE Verify $n_{q1}$ and $n_{q2}$
\STATE $unsafe\_circuit \gets random\_qc$ without $safe\_qubit$
\STATE $ops\_removed \gets 0$
\WHILE{$n_{q1} > 0$ \OR $n_{q2} > 0$}
    \STATE $remove\_loc \gets$ random int between $0$ and $(unsafe\_circuit\ length -1)$
    \STATE $op \gets$ gate in $unsafe\_circuit$ at index $remove\_loc$
    \STATE $gate\_to\_remove \gets None$
    \IF{$op$ input qubits does not contain $safe\_qubit$}
        \IF{number of $op$ input qubits = 1 \AND $n_{q1} > 0$}
            \STATE $gate\_to\_remove \gets op$
            \STATE Decrement $n_{q1}$ by 1
            \STATE Increment $ops\_removed$ by 1
        \ELSIF{number of $op$ input qubits = 2 \AND $n_{q2} > 0$}
            \STATE $gate\_to\_remove \gets op$
            \STATE Decrement $n_{q2}$ by 1
            \STATE Increment $ops\_removed$ by 2
        \ENDIF
        \IF{$gate\_to\_remove \neq \text{None}$}
            \STATE Remove $gate\_to\_remove$ from $random\_qc.data$
            \STATE Remove $gate\_to\_remove$ from $unsafe\_circuit$
        \ENDIF
    \ENDIF
\ENDWHILE
\STATE Check $ops\_removed = ops\_to\_remove$
\STATE \textbf{Output:} random\_qc
\end{algorithmic}
\end{algorithm}

This algorithm facilitates the generation of random circuits with controlled depth, width and gate density. This enables a large-scale dataset of circuits to be produced to test problems at any desired scale, allowing the future scale of quantum algorithms to be considered for novel concepts. 

The effectiveness of this random circuit generator in supporting testing of accelerated compilers for large-scale circuits is described in the Results section.

\subsection{Parallel Compilation}\label{sec:parallel_methodology}
\begin{figure}[tbp]
    \centering
    \includegraphics[width=0.5\linewidth]{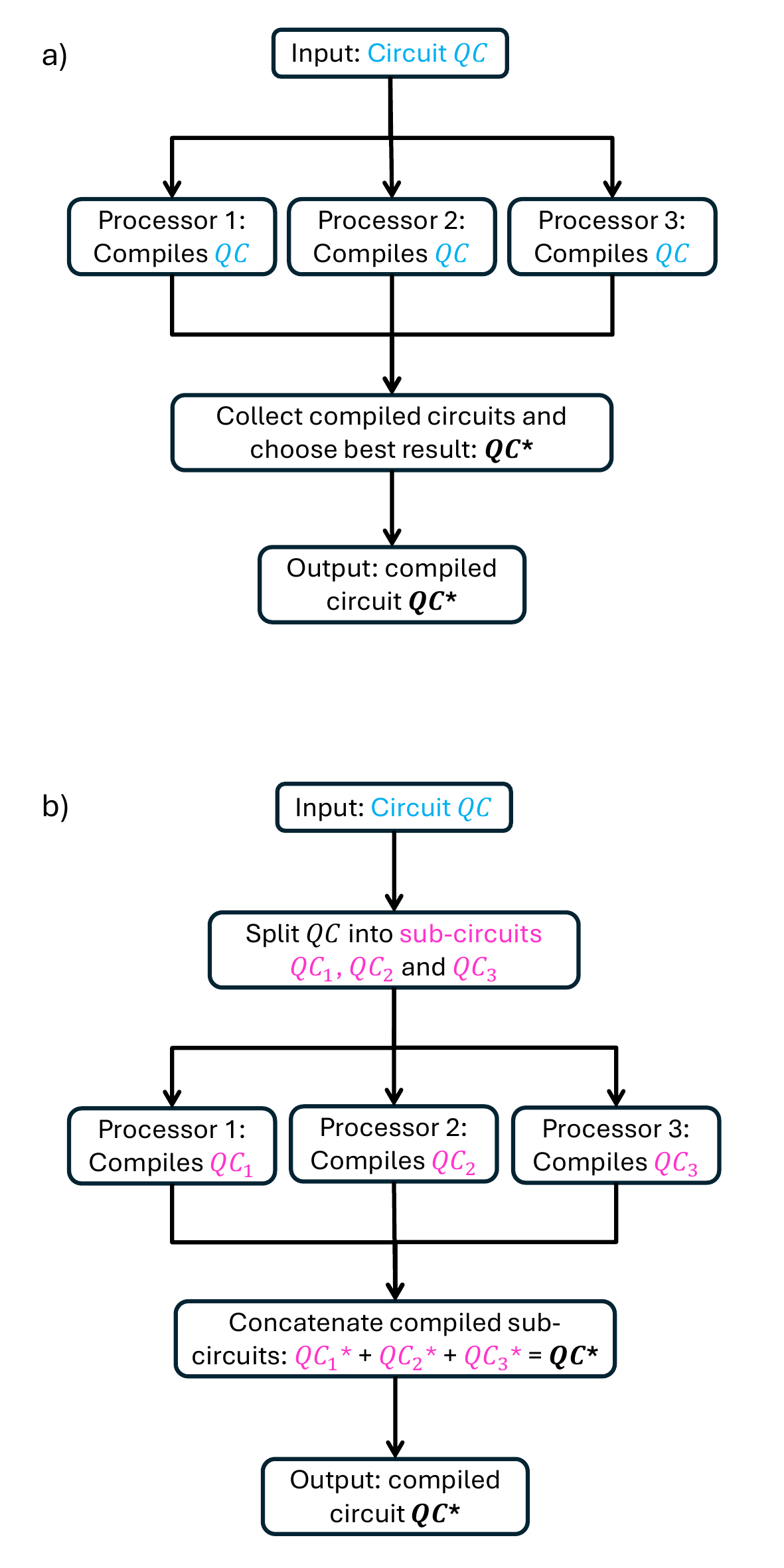}
    \caption{Flow charts illustrating the workflows for a) Qiskit and b) the parallel compilation approach. Both use 3 processors.}
    \label{fig:parallel_comparision}
\end{figure}

Circuit compilation is an NP-Complete problem which leads to long compilation times \cite{Qubit_allocation} \cite{compilation_complexity}, superseding the circuit execution time for large circuits \cite{Nation2025}. The compilation pipeline used by Qiskit's SabreSwap algorithm is illustrated in Fig. \ref{fig:parallel_comparision}a). Each processor receives a copy of the original quantum circuit and independently compiles it given the quantum lattice limitations. Upon task completion across all processors, the best circuit is chosen i.e. the circuit with the least SWAP gates inserted \cite{SABRE_PROP}. This paper proposes decomposing the original circuit into sub-circuits, which are compiled in parallel on independent cores. The compiled sub-circuits are concatenated to produce the final program. This is illustrated in Fig. \ref{fig:parallel_comparision}b). In comparison with SabreSwap this allocates only a portion of the original circuit to each independent processor to be compiled in parallel. This approach should reduce compilation time while maintaining circuit quality, quantified by the depth, gate and SWAP count.

\begin{figure}[tbp]
  \centering
  \begin{minipage}[b]{0.6\linewidth}
    \centering
    \includegraphics[width=1.0\linewidth, trim={0 0 0 1.2cm},clip]{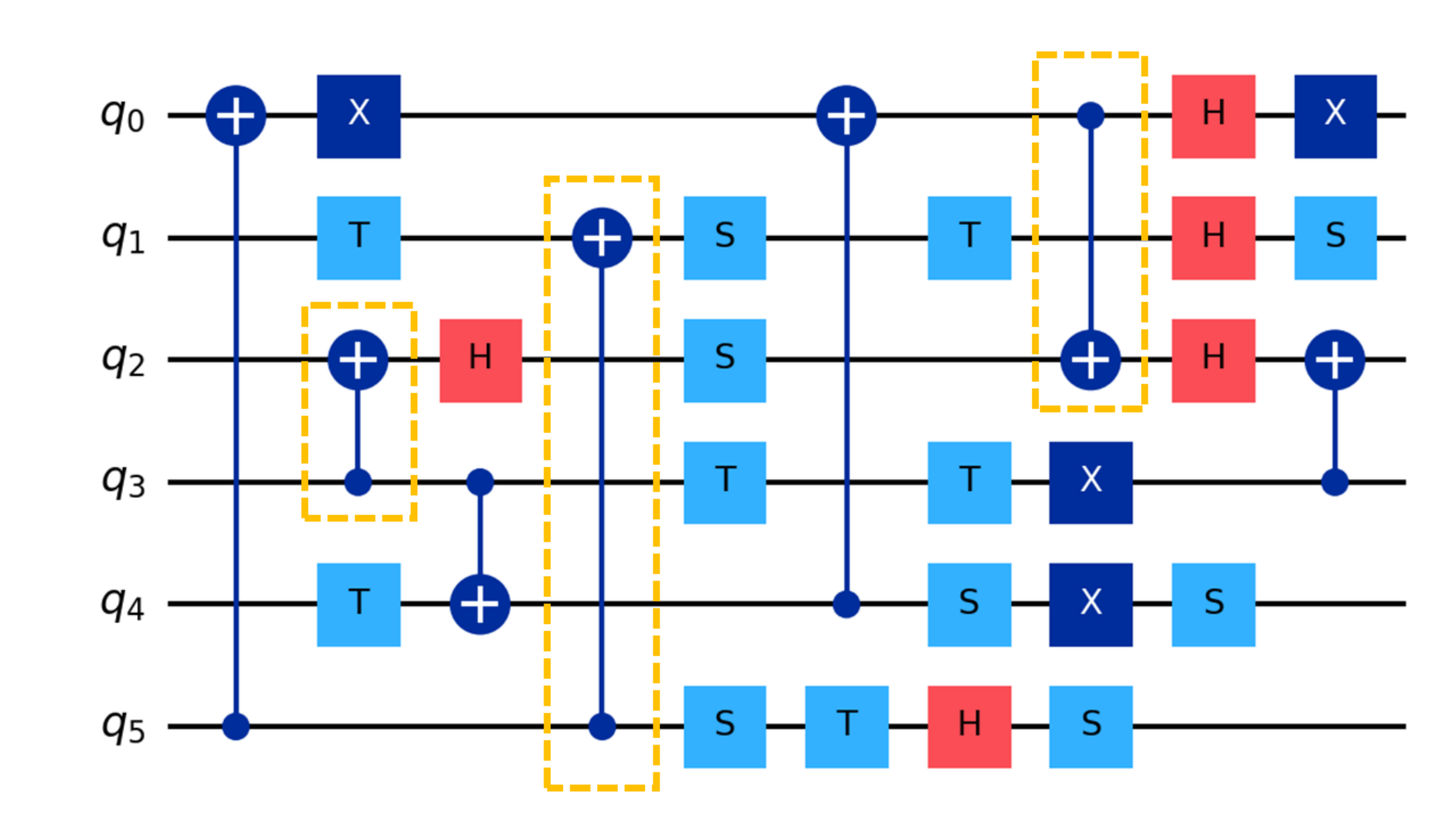}
    \caption{A randomly-generated six-qubit circuit, of depth 6 gates. Boxes denote 1-qubit quantum gates and time flows left to right. The box colours are agonistic of meaning. This circuit will be compiled to run on a six-qubit, $2 \times 3$ grid processor. Examples of 2-qubit (CNOT) gates which cannot be executed on the processor as is have been outlined in orange.}
    \label{fig:og-circuit}
  \end{minipage}
  \hfill
  \begin{minipage}[b]{0.35\linewidth}
    \centering
    \includegraphics[width=1.0\linewidth, trim={0 0cm 0 0cm},clip]{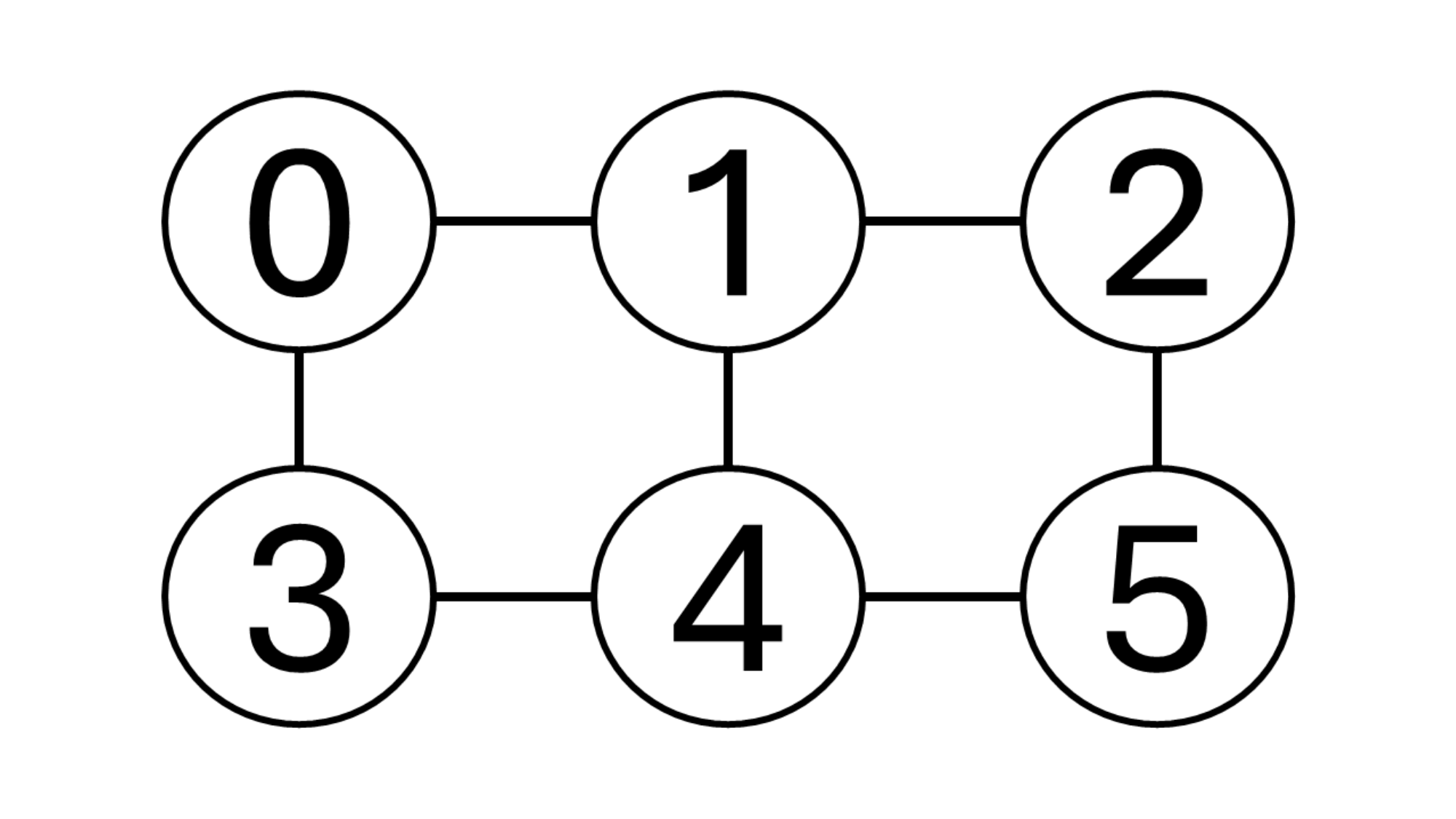}
    \vspace{5mm}
    \caption{6-Qubit Processor Layout. The qubits are arranged in a $2 \times 3$ grid layout. The circles represent qubits and the lines are physical connections.}
    \label{fig:six-qubit-proc}
  \end{minipage}
\end{figure}

An illustrative example circuit is shown in Fig. \ref{fig:og-circuit}. This is a trivial example circuit, the techniques proposed in this work are aimed at large-scale circuits with hundreds of qubits and millions of gates. This circuit will be compiled for the example processor layout illustrated in Fig. \ref{fig:six-qubit-proc} which mimics the ladder layout of IBM Melbourne \cite{ibm_melbourne_device}. Considering NNA, three examples of CNOT gates that cannot be executed on the given processor are outlined in orange in Fig. \ref{fig:og-circuit}. To enable these CNOT gates to operate on neighbouring qubits, SWAP gates will be inserted during compilation to adjust the physical qubit mapping.

First the original circuit is decomposed into a series of sub-circuits which are then compiled in parallel on independent processors. The number of sub-circuits is equal to the number of cores. The number of gates in each sub-circuit $g_{sc}$ is determined as in (\ref{eq:sc_gates}), where $n_{g}$ is the number of gates in the original circuit and $n_{sc}$ is the number of sub-circuits. 
\begin{equation} \label{eq:sc_gates}
g_{sc} = \left\lfloor \frac{n_{g}}{n_{sc}} \right\rfloor
\end{equation}

The list of circuit gate instructions is split into $n_{sc}$ chunks of length $g_{sc}$, any remainder gates are allocated to the final chunk. %, preceded by the header. 
The list of gate operations is parsed directly since this is the actual input to a compiler (in the form of, for example, QASM) and early experiments revealed a significant time overhead creating a quantum circuit representation after loading the instruction list. Consequently, circuit decomposition is handled naïvely by separating the instruction list.

\begin{figure}[tbp]
    \centering
    \includegraphics[width=0.75\linewidth, trim={0cm 2cm 0 2cm},clip]{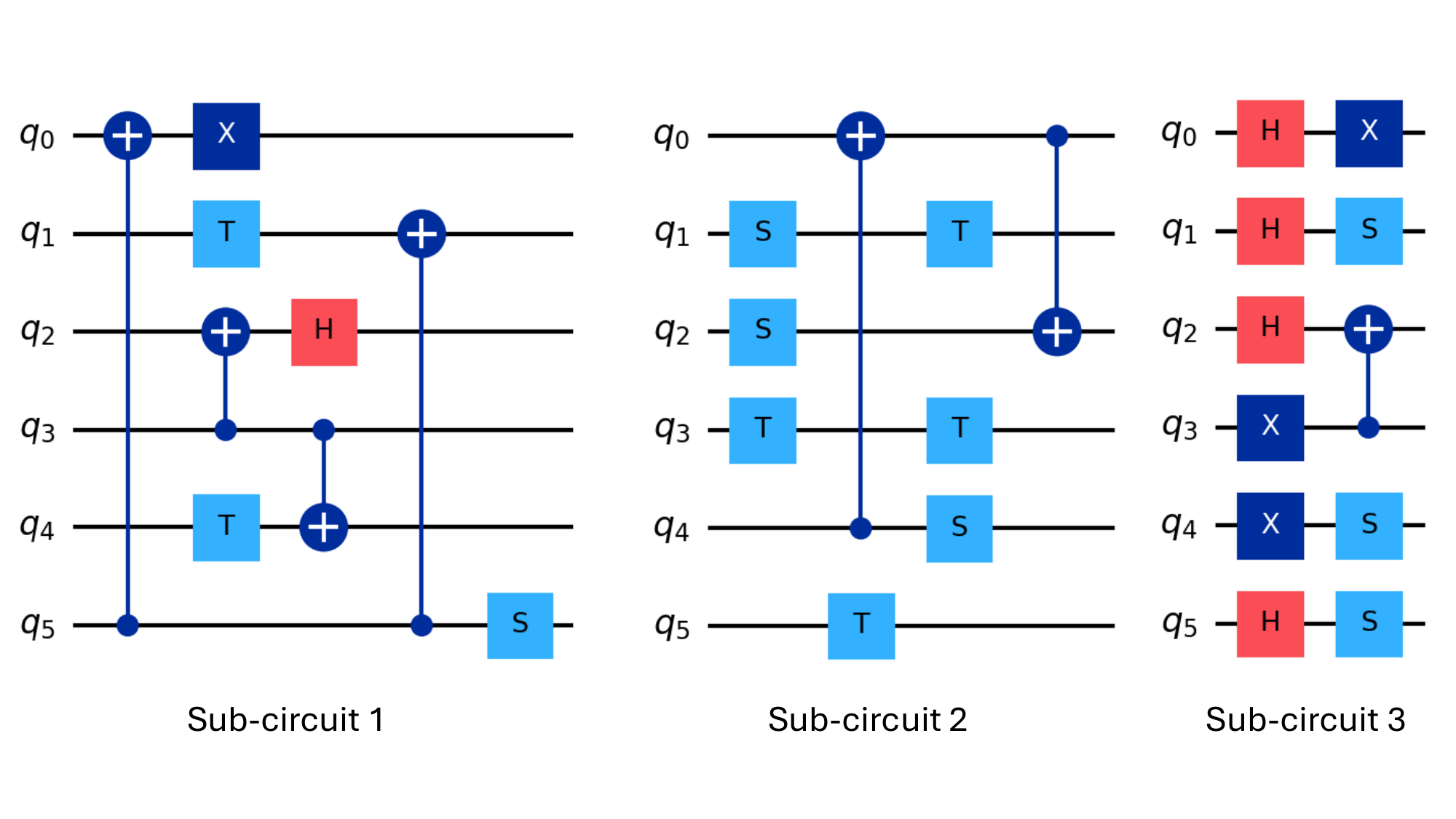}
    \vspace{-3mm}
    \caption{The sub-circuits produced from the decomposition of the example circuit. Sub-circuits 1 and 2 have 9 gates. Sub-circuit 3 has 11 gates. Note the initial qubit orderings are all trivial. As before, the boxes denote 1-qubit quantum gates, solid dots are control qubits, and $\oplus$ are target qubits of 2-qubit CNOT gates. The box colours are used for visual only, with no additional meaning.}
    \label{fig:sub-circuits}
\end{figure}

The example circuit in Fig. \ref{fig:og-circuit} is separated into three sub-circuits, which can be seen in Fig. \ref{fig:sub-circuits}. There are 29 gates in the original circuit, hence according to (\ref{eq:sc_gates}), the first 2 sub-circuits consist of 9 gates, the 2 remainder gates are placed in the final sub-circuit, giving it 11 gates. This balances the computational load as the final sub-circuit needs less post-processing following compilation.

\begin{figure}[htbp]
  \centering
  \begin{minipage}[b]{0.48\linewidth}
    \centering
    \includegraphics[width=1.0\linewidth, trim={5cm 2cm 4cm 3cm},clip]{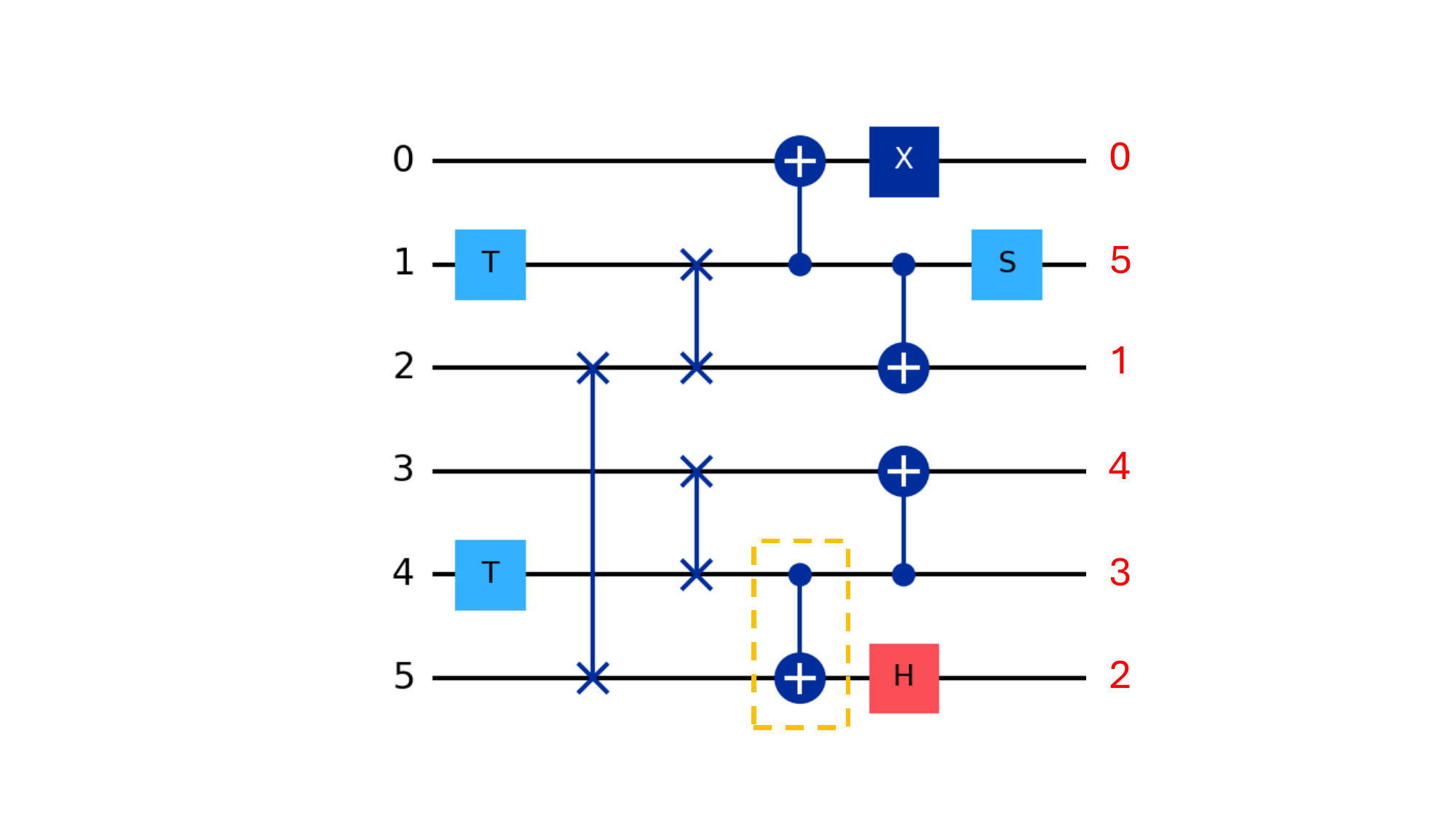}
    \caption{Sub-circuit 1 compiled using Qiskit SabreSwap. Again boxes denote 1-qubit gates, solid dots and $\oplus$ are CNOT gates and colours are for visual clarity only. The CNOT in the orange box corresponds to the first CNOT outlined in Fig. \ref{fig:og-circuit}. SWAP gates are shown as 2 $\times$ symbols connected by a vertical line. The insertion of the SWAP gates has altered the logical-to-physical qubit mapping throughout the duration of the circuit. Hence the output qubit order denoted in red does not match the input order on the left.}
    \label{fig:sub-circuit-1-compiled-no_pc}
  \end{minipage}
  \hfill
  \begin{minipage}[b]{0.50\linewidth}
    \centering
    \includegraphics[width=1.0\linewidth, trim={2cm 0cm 0.5cm, 0.5cm},clip]{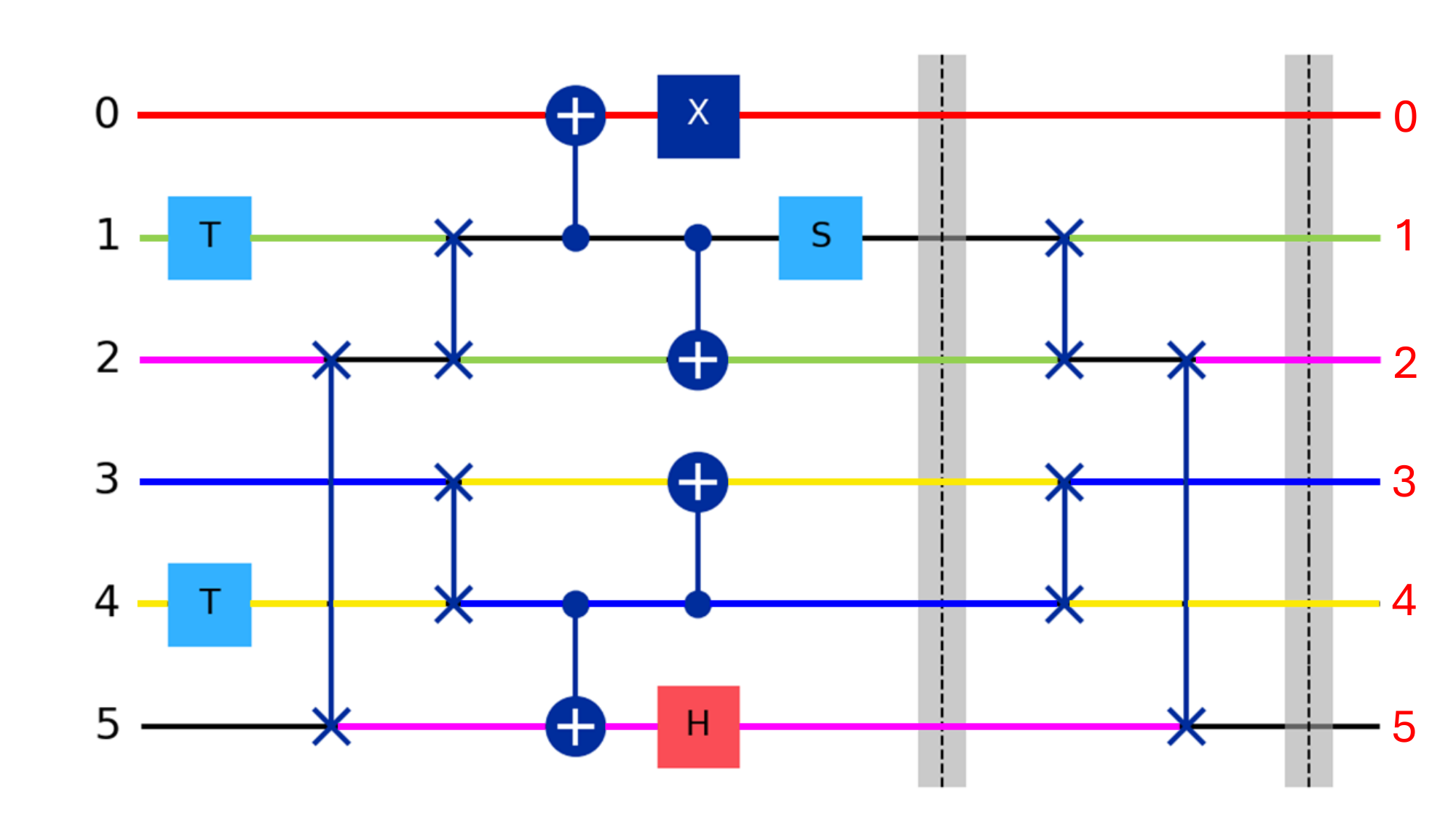}
    \caption{Compiled example sub-circuit 1 with permutation circuit appended. The colours correspond to the physical position of the logical qubits i.e. magenta follows the logical state of qubit 2 as it is moved on the physical processor by the SWAP gates. Note time runs left to right. The vertical grey lines denote barriers, which are only used to visually separate the circuit. The permutation circuit is contained within the barriers. As a result of the permutation circuit, the ordering of the input (left) and output (right) qubits match. }
    \label{fig:sub-circuit-1-compiled_colour}
  \end{minipage}
\end{figure}

Following decomposition the sub-circuits are compiled in parallel on independent cores. 
Two different compilers have been explored in this work: the Qiskit Transpiler and PyTKET. Fig. \ref{fig:sub-circuit-1-compiled-no_pc} shows example sub-circuit 1 after it has been compiled with Qiskit for the $2 \times 3$ processor in Fig. \ref{fig:six-qubit-proc}. Three SWAP gates are inserted to facilitate NNA. The SWAP gates move the logical states on the physical qubits e.g, the first operation on qubit 2 is a SWAP gate which swaps the logical qubit states on qubits 2 and 5. The CNOT highlighted in the orange box, operating on qubits 4 and 5, corresponds to the first CNOT outlined in orange in Fig. \ref{fig:og-circuit}. Qubits 4 and 5 are neighbours on the processor and hence the CNOT can now be executed. 

Prior to compilation the depth of sub-circuit 1 was 3 gates deep, with 9 gates total. Post compilation the depth is 5 gates deep with a total gate count of 12 gates. Hence, the compilation process changes the physical characteristics of the circuit. Typically compilation algorithms will aim to minimise the number of gates and the depth as increasing these characteristics increases the noise introduced during circuit execution \cite{QAOA_compilation} \cite{synth_time_optimal}. Hence for the parallel approach it is important that the parallelisation does not cause extreme overheads in the number of gates or the depth.

The initial order of the logical qubits is $[0, 1, 2, 3, 4, 5]$ for the sub-circuit in Fig. \ref{fig:sub-circuit-1-compiled-no_pc}. This is the case by default for all of the sub-circuits produced from the original circuit. Due to the insertion of the SWAP gates, the final logical qubit order is $[0, 5, 1, 4, 3, 2]$. Due to this re-arrangement, sub-circuits cannot simply be concatenated back together as this would result in gates operating on the incorrect quantum states, altering the behaviour of the circuit. Hence, a method is needed to reset the qubit mapping at the end of the sub-circuit.

A permutation circuit transforms one qubit mapping into another \cite{leo_thesis}. A permutation circuit only consists of SWAP gates and is used to reorder the qubits states into the desired logical-to-physical qubit mapping. % The permutation circuit is appended to a sub-circuit.
The A* Algorithm is used on the final qubit ordering to determine the path lengths from the current to targeted qubit locations. The shortest path is taken and the SWAP gates are implemented to move the qubit state into the correct position. This is repeated until all qubits are correctly mapped. The permutation circuit is appended to the sub-circuit.

Algorithm \ref{alg:perm_circ} implements permutation circuit generation: 
\begin{itemize}
    \item The inputs: $final\_layout$ is a list of qubits in the final positions, $width$ is the number of qubits, $compiled\_subcirc$ is the compiled sub-circuit and $c\_map$ is the processor connectivity graph.
    \item Lines 2-4 establish the initial layout and the stopping condition that ensures every qubit is in the targeted location. 
    \item Lines 5-10 iterate over each qubit in the initial layout and computes the distance between the current and target locations using the A* algorithm and the processor connectivity graph. The distance and path information is stored for each qubit.
    \item Lines 11-20 find the qubit closest to it's targeted location. The path for this qubit is retrieved and walked along, adding a SWAP between each qubit pair in the path to the swap list. The final mapping is updated corresponding to the swapped qubits.
    \item Lines 21-28 calculate the updated distance between the current and desired locations using the updated final qubit mapping. This loops back to line 11 until every qubit is in the targeted location and the stopping condition is fulfilled i.e. all the path lengths are 1.
    \item Lines 30-31 apply the SWAP operations in the swap list on the compiled sub-circuit.
\end{itemize}

Fig. \ref{fig:sub-circuit-1-compiled_colour} illustrates the movement of the logical qubit states throughout the duration of sub-circuit 1. The vertical gray lines denote the start and end of the permutation circuit. 
The colours represent the physical positions of the logical qubits. Because the colour ordering at the start of the sub-circuit matches the ordering at the end of the permutation circuit, the final qubit mapping is as required. This enables the second sub-circuit to be appended directly to the first, as the final mapping of the first sub-circuit aligns with the initial mapping of the second.

Once again the depth and number of gates in sub-circuit 1 has been altered, the depth of the sub-circuit is now 7 gates deep with 15 gates total. It can be seen that the permutation circuit impacts the circuit gate count and depth. Specifically, permutation circuits may introduce more SWAP gates into the circuit than would be inserted if the original circuit was compiled as one. The SWAP overhead should be considered alongside the depth and gate count overheads when using this process to compile circuits.

\begin{figure}[htbp]
    \centering
    \includegraphics[width=0.95\linewidth, trim={0 3cm 0 6cm},clip]{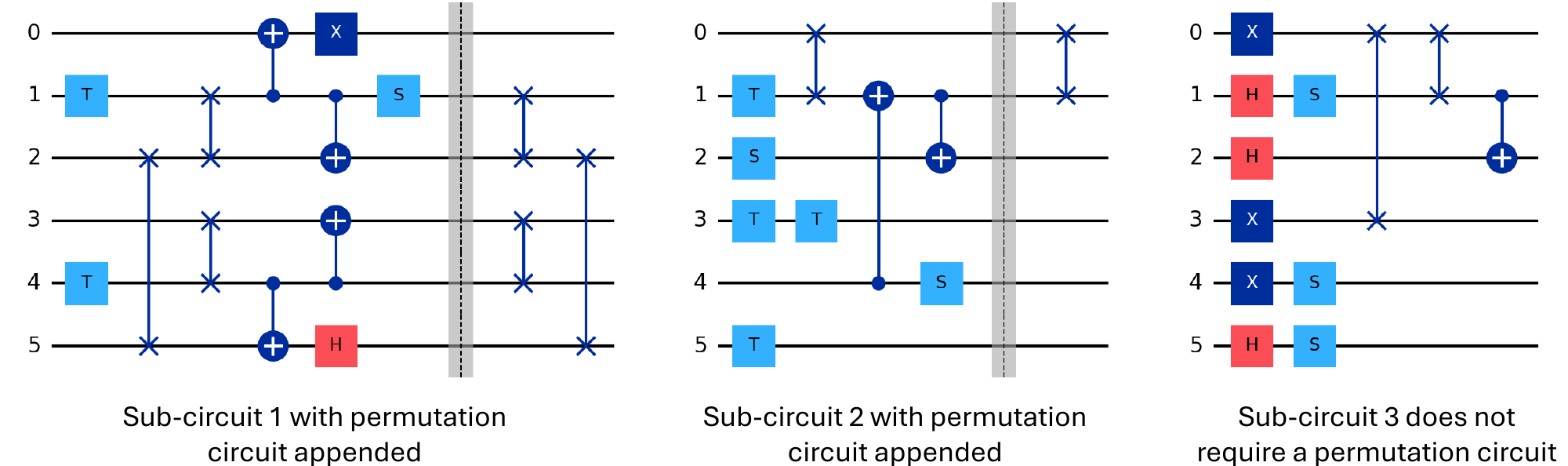}
    \vspace{-3mm}
    \caption{Compiled sub-circuits with permutation circuits appended. As before time runs left to right for each sub-circuit. The boxes denote 1-qubit gates with colours for visualisation only. CNOT gates are represented by control (large dots) and target ($\oplus$) qubits. SWAP gates are $\times$ symbols connected by vertical lines. The vertical grey lines are the barriers which denote the insertion point for each permutation circuits.}
    \label{fig:sub-circuits_with_perm}
\end{figure}

Fig. \ref{fig:sub-circuits_with_perm} shows all 3 compiled example sub-circuits, with the permutation circuits appended to sub-circuits 1 and 2. The final sub-circuit does not require a permutation circuit as the final logical-to-physical qubit mapping is not of importance. This is the reason the load balancing allocates any remainder gates to the final sub-circuit during circuit decomposition. The final sub-circuit can accommodate a larger section of the original circuit, as it does not require additional time for the permutation circuit generation.

\begin{figure}[bp]
    \centering
    \includegraphics[width=0.95\linewidth, trim={0 5cm 0 5cm},clip]{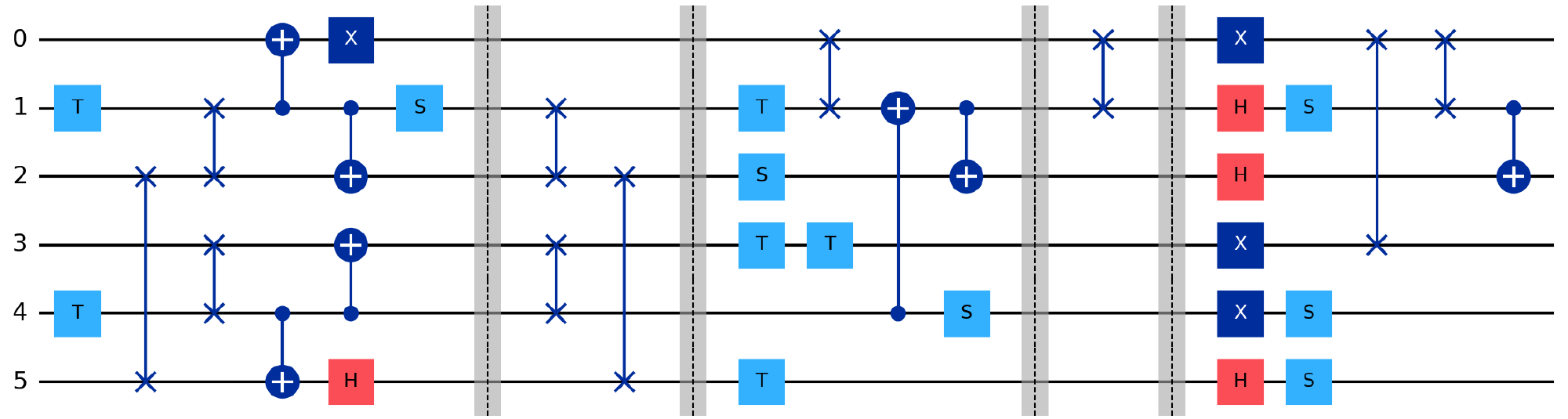}
    \vspace{-3mm}
    \caption{Final compiled circuit. Time runs left to right. The boxes denote 1-qubit gates (colour-agnostic). Large dots and $\oplus$ connected by a line denote CNOT gates. SWAP gates are $\times$ symbols connected by vertical lines. The permutation circuits are bounded by barriers (gray vertical lines).}
    \label{fig:final-compiled-circuit}
\end{figure}

The compiled sub-circuits are concatenated in order to produce a final circuit. Again this is handled by parsing the sub-circuit instructions lists directly to prevent the overheads of loading the quantum circuit objects. Fig. \ref{fig:final-compiled-circuit} illustrates the final compiled circuit which can now be executed on the processor shown in Fig. \ref{fig:six-qubit-proc} as it is NNA compliant. The final compiled circuit depth is 16 gates deep. The original circuit was only 6 gates deep. This example circuit is only a very short illustrative example, but clearly the overheads must be considered when using the parallel compilation process. When the Qiskit Transpiler was used to compile original circuit in Fig. \ref{fig:og-circuit} as a whole, only 5 SWAP gates were required and the final post-compilation depth was 10 gates. Hence, the parallel compilation approach should not be considered for trivially short circuits.

\begin{algorithm}
\caption{Generating a Permutation Circuit}
\begin{algorithmic}[1]\label{alg:perm_circ}
\STATE \textbf{Input:} $final\_layout$, $width$, $compiled\_subcirc$, $c\_map$
\STATE $initial\_layout \gets$ \text{list}(0, 1, \ldots, $width$ - 1)

\STATE $sorted\_layout \gets [1, 1, \ldots, 1]$ \COMMENT{List of length $width$}
\STATE $dists$, $paths$, $swap\_list \gets$ empty list 
% get_dists
\FOR{$qubit$ in $initial\_layout$}
    \STATE $pos \gets$ index of $qubit$ in $final\_layout$
    \STATE $dist \gets$ path from $qubit$ to $pos$ on the $c\_map$ using the A* algorithm
    \STATE Append $dist$ to $paths$
    \STATE Append $dist.length()$ to $dists$
\ENDFOR

\WHILE{$dists \neq sorted\_layout$}
    % \STATE $index \gets \text{index\_of}(\textbf{min}\{i \in dists \;|\; i > 1\})$
    \STATE $path \gets $ path at index of $\textbf{min}\{i \in dists \;|\; i > 1\})$
    \FOR{$counter = 0$ \TO $path.length()-1$}
        \STATE $i \gets path[counter]$
        \STATE $j \gets path[counter + 1]$
        \STATE $tmp \gets final\_layout[i]$
        \STATE $final\_layout[i] \gets final\_layout[j]$
        \STATE $final\_layout[j] \gets tmp$
        \STATE Append $i$, $j$ pair to $swap\_list$
    \ENDFOR
    \STATE $dists$, $paths \gets$ empty list
    \FOR{$qubit$ in $initial\_layout$}
        \STATE $pos \gets$ index of $qubit$ in $final\_layout$
        \STATE $dist \gets$ path from $qubit$ to $pos$ on the $c\_map$ using the A* algorithm
        \STATE Append $dist$ to $paths$
        \STATE Append $dist.length()$ to $dists$
    \ENDFOR
\ENDWHILE

\FOR{$i$, $j$ in $swap\_list$}
    \STATE Apply a SWAP on qubits $i$ and $j$ in $compiled\_subcirc$
\ENDFOR
\STATE \textbf{Output:} $compiled\_sc$
\end{algorithmic}
\end{algorithm}

\newpage
\bibliographystyle{naturemag}
\bibliography{references}

@INPROCEEDINGS{NNA,
  author={Y. {Hirata} and M. {Nakanishi} and S. {Yamashita} and  Y. {Nakashima}},
  booktitle={2009 Third International Conference on Quantum, Nano and Micro Technologies}, 
  title={An Efficient Method to Convert Arbitrary Quantum Circuits to Ones on a Linear Nearest Neighbor Architecture}, 
  year={February 2009},
  volume={},
  number={},
  pages={},
  doi={10.1109/ICQNM.2009.25},
  url={https://doi.org/10.1109/ICQNM.2009.25}
}

@misc{qc_with_qiskit,
      title={Quantum computing with Qiskit}, 
      author={Ali Javadi-Abhari and Matthew Treinish and Kevin Krsulich and Christopher J. Wood and Jake Lishman and Julien Gacon and Simon Martiel and Paul D. Nation and Lev S. Bishop and Andrew W. Cross and Blake R. Johnson and Jay M. Gambetta},
      year={2024},
      eprint={2405.08810},
      archivePrefix={arXiv},
      primaryClass={quant-ph},
      url={https://arxiv.org/abs/2405.08810}, 
}

@phdthesis{leo_thesis,
    title    = {The synthesis of nearest neighbour compliant quantum circuits},
    school   = {Queen's University Belfast},
    author   = {L. {Rogers}},
    year     = {2021},
    url      = {https://pure.qub.ac.uk/en/studentTheses/the-synthesis-of-nearest-neighbour-compliant-quantum-circuits}
}

@INPROCEEDINGS{SABRE_PROP,
  author={G. {Li} and Y. {Ding} and Y. {Xie}},
  title={Tackling the Qubit Mapping Problem for NISQ-Era Quantum Devices}, 
  booktitle={ASPLOS '19: Proceedings of the Twenty-Fourth International Conference on Architectural Support for Programming Languages and Operating Systems}, 
  year={2019},
  pages = {1001-–1014},
  url = {https://doi.org/10.1145/3297858.3304023}
}

@INPROCEEDINGS{11009047,
  author={Hai, Vu Tuan and Duong, Le Vu Trung and Luan, Pham Hoai and Nakashima, Yasuhiko},
  booktitle={2024 RIVF International Conference on Computing and Communication Technologies (RIVF)}, 
  title={Efficient Random Quantum Circuit Generator: A Benchmarking Approach for Quantum Simulators}, 
  year={2024},
  volume={},
  number={},
  pages={419-423},
  doi={10.1109/RIVF64335.2024.11009047}}

@article{Wagner2023,
  author    = {Florian Wagner and Armin B{\"a}rmann and Frauke Liers and others},
  title     = {Improving Quantum Computation by Optimized Qubit Routing},
  journal   = {Journal of Optimization Theory and Applications},
  volume    = {197},
  pages     = {1161--1194},
  year      = {2023},
  doi       = {10.1007/s10957-023-02229-w}
}

@inproceedings{qmap,
  title = {{MQT QMAP: Efficient Quantum Circuit Mapping}},
  booktitle = {International Symp. on Physical Design},
  author = {Wille, Robert and Burgholzer, Lukas},
  year = {2023},
  doi = {10.1145/3569052.3578928},
}

@article{mqtbench,
  title={{{MQT Bench}}: Benchmarking Software and Design Automation Tools for Quantum Computing},
  shorttitle = {{MQT Bench}},
  journal = {{Quantum}},
  author={Quetschlich, Nils and Burgholzer, Lukas and Wille, Robert},
  year={2023},
  note={{{MQT Bench}} is available at \url{https://www.cda.cit.tum.de/mqtbench/}},
}

@article{qasmbench,
author = {Li, Ang and Stein, Samuel and Krishnamoorthy, Sriram and Ang, James},
title = {QASMBench: A Low-Level Quantum Benchmark Suite for NISQ Evaluation and Simulation},
year = {2023},
issue_date = {June 2023},
publisher = {Association for Computing Machinery},
address = {New York, NY, USA},
volume = {4},
number = {2},
url = {https://doi.org/10.1145/3550488},
doi = {10.1145/3550488},
journal = {ACM Transactions on Quantum Computing},
month = feb,
articleno = {10},
numpages = {26}
}

@misc{red-queen,
  author       = {{IBM Qiskit}},
  title        = {Red Queen: Quantum software benchmarking tool},
  year         = {2024},
  url          = {https://github.com/Qiskit/red-queen},
  note         = {Archived on August 14, 2024}
}

@article{parallel_paper,
  title={A Parallelized Qubit Mapping Algorithm for Large-scale Quantum Circuits},
  author={Dongmin Kim and Sengthai Heng and Youngsun Han},
  journal={IEIE SPC(IEIE Transactions on Smart Processing and Computing)},
  volume={11},
  number={01},
  pages={40-48},
  year={2022},
  doi={https://doi.org/10.5573/IEIESPC.2021.11.1.40},
  issn={2287-5255}
}

@article{synth_time_optimal,
  doi = {10.22331/q-2023-04-20-984},
  url = {https://doi.org/10.22331/q-2023-04-20-984},
  title = {Synthesis of and compilation with time-optimal multi-qubit gates},
  author = {Ba{\ss{}}ler, Pascal and Zipper, Matthias and Cedzich, Christopher and Heinrich, Markus and Huber, Patrick H. and Johanning, Michael and Kliesch, Martin},
  journal = {{Quantum}},
  issn = {2521-327X},
  publisher = {{Verein zur F{\"{o}}rderung des Open Access Publizierens in den Quantenwissenschaften}},
  volume = {7},
  pages = {984},
  month = apr,
  year = {2023}
}

@inproceedings{Qubit_allocation,
author = {Siraichi, Marcos Yukio and Santos, Vin\'{\i}cius Fernandes dos and Collange, Caroline and Pereira, Fernando Magno Quintao},
title = {Qubit allocation},
year = {2018},
isbn = {9781450356176},
publisher = {Association for Computing Machinery},
address = {New York, NY, USA},
url = {https://doi.org/10.1145/3168822},
doi = {10.1145/3168822},
booktitle = {Proceedings of the 2018 International Symposium on Code Generation and Optimization},
pages = {113–125},
numpages = {13},
location = {Vienna, Austria},
series = {CGO '18}
}

@inproceedings{compilation_complexity,
  title     = {On the Complexity of Quantum Circuit Compilation},
  author    = {Adi Botea and Akihiro Kishimoto and Radu Marinescu},
  booktitle = {Proceedings of the Eleventh Annual Symposium on Combinatorial Search (SoCS)},
  year      = {2018},
  pages     = {138--142},
  publisher = {AAAI Press},
  doi       = {10.1609/socs.v11i1.18463},
  url       = {https://ojs.aaai.org/index.php/SOCS/article/view/18463}
}

@article{factor2048,
  doi = {10.22331/q-2021-04-15-433},
  url = {https://doi.org/10.22331/q-2021-04-15-433},
  title = {How to factor 2048 bit {RSA} integers in 8 hours using 20 million noisy qubits},
  author = {Gidney, Craig and Eker{\aa{}}, Martin},
  journal = {{Quantum}},
  issn = {2521-327X},
  publisher = {{Verein zur F{\"{o}}rderung des Open Access Publizierens in den Quantenwissenschaften}},
  volume = {5},
  pages = {433},
  month = apr,
  year = {2021}
}

@article{Muller2024,
  author = {Müller, S. and Phillipson, F.},
  title = {Quantum annealing for nearest neighbour compliance problem},
  journal = {Scientific Reports},
  volume = {14},
  pages = {23340},
  year = {2024},
  publisher = {Nature Publishing Group},
  doi = {10.1038/s41598-024-73882-y},
  url = {https://doi.org/10.1038/s41598-024-73882-y}
}

@ARTICLE{reordering,
  author={Wille, Robert and Lye, Aaron and Drechsler, Rolf},
  journal={IEEE Transactions on Computer-Aided Design of Integrated Circuits and Systems}, 
  title={Exact Reordering of Circuit Lines for Nearest Neighbor Quantum Architectures}, 
  year={2014},
  volume={33},
  number={12},
  pages={1818-1831},
  doi={10.1109/TCAD.2014.2356463}
}

@inproceedings{pytket_qubit_routing,
  doi = {10.4230/LIPICS.TQC.2019.5},
  url = {https://drops.dagstuhl.de/entities/document/10.4230/LIPIcs.TQC.2019.5},
  author = {Cowtan, Alexander and Dilkes, Silas and Duncan, Ross and Krajenbrink, Alexandre and Simmons, Will and Sivarajah, Seyon},
  language = {en},
  title = {On the Qubit Routing Problem},
  publisher = {Schloss Dagstuhl – Leibniz-Zentrum für Informatik},
  year = {2019},
  copyright = {Creative Commons Attribution 3.0 Unported license}
}

@INPROCEEDINGS{QAOA_compilation,
  author={Alam, Mahabubul and Ash-Saki, Abdullah and Ghosh, Swaroop},
  booktitle={2020 53rd Annual IEEE/ACM International Symposium on Microarchitecture (MICRO)}, 
  title={Circuit Compilation Methodologies for Quantum Approximate Optimization Algorithm}, 
  year={2020},
  volume={},
  number={},
  pages={215-228},
  doi={10.1109/MICRO50266.2020.00029}}

@INPROCEEDINGS{correct_compilation,
  author={Wille, Robert and Hillmich, Stefan and Burgholzer, Lukas},
  booktitle={2020 IEEE International Symposium on Circuits and Systems (ISCAS)}, 
  title={Efficient and Correct Compilation of Quantum Circuits}, 
  year={2020},
  volume={},
  number={},
  pages={1-5},
  doi={10.1109/ISCAS45731.2020.9180791}}

@article{Venturelli_2018,
doi = {10.1088/2058-9565/aaa331},
url = {https://dx.doi.org/10.1088/2058-9565/aaa331},
year = {2018},
month = {feb},
publisher = {IOP Publishing},
volume = {3},
number = {2},
pages = {025004},
author = {Venturelli, Davide and Do, Minh and Rieffel, Eleanor and Frank, Jeremy},
title = {Compiling quantum circuits to realistic hardware architectures using temporal planners},
journal = {Quantum Science and Technology}
}

@INPROCEEDINGS{wille_compilation_time,
  author={Quetschlich, Nils and Burgholzer, Lukas and Wille, Robert},
  booktitle={2023 IEEE International Conference on Quantum Computing and Engineering (QCE)}, 
  title={Reducing the Compilation Time of Quantum Circuits Using Pre-Compilation on the Gate Level}, 
  year={2023},
  volume={01},
  number={},
  pages={757-767},
  doi={10.1109/QCE57702.2023.00091}}

@article{PhysRevA.70.052328,
  title = {Improved simulation of stabilizer circuits},
  author = {Aaronson, Scott and Gottesman, Daniel},
  journal = {Phys. Rev. A},
  volume = {70},
  issue = {5},
  pages = {052328},
  numpages = {14},
  year = {2004},
  month = {Nov},
  publisher = {American Physical Society},
  doi = {10.1103/PhysRevA.70.052328},
  url = {https://link.aps.org/doi/10.1103/PhysRevA.70.052328}
}

@article{gambetta_future,
    author = {Bravyi, Sergey and Dial, Oliver and Gambetta, Jay M. and Gil, Darío and Nazario, Zaira},
    title = {The future of quantum computing with superconducting qubits},
    journal = {Journal of Applied Physics},
    volume = {132},
    number = {16},
    pages = {160902},
    year = {2022},
    month = {10},
    issn = {0021-8979},
    doi = {10.1063/5.0082975},
    url = {https://doi.org/10.1063/5.0082975},
    eprint = {https://pubs.aip.org/aip/jap/article-pdf/doi/10.1063/5.0082975/20034201/160902_1_5.0082975.pdf},
}

@article{Nation2025,
  author    = {Nation, P. D. and Saki, A. A. and Brandhofer, S. and others},
  title     = {Benchmarking the performance of quantum computing software for quantum circuit creation, manipulation and compilation},
  journal   = {Nature Computational Science},
  year      = {2025},
  volume    = {5},
  pages     = {427--435},
  doi       = {10.1038/s43588-025-00792-y},
  url       = {https://doi.org/10.1038/s43588-025-00792-y}
}

@phdthesis{feynman_suite,
  author       = {Amy, Matthew},
  title        = {Formal Methods in Quantum Circuit Design},
  school       = {University of Waterloo},
  address      = {Waterloo, Ontario, Canada},
  year         = {2019},
  type         = {PhD thesis},
  url          = {http://hdl.handle.net/10012/14480},
  note         = {Advisor: Michele Mosca}
}

@misc{veri_q,
      title={VeriQBench: A Benchmark for Multiple Types of Quantum Circuits}, 
      author={Kean Chen and Wang Fang and Ji Guan and Xin Hong and Mingyu Huang and Junyi Liu and Qisheng Wang and Mingsheng Ying},
      year={2022},
      eprint={2206.10880},
      archivePrefix={arXiv},
      primaryClass={quant-ph},
      url={https://arxiv.org/abs/2206.10880}, 
}

@INPROCEEDINGS {supermarq,
author = { Tomesh, Teague and Gokhale, Pranav and Omole, Victory and Ravi, Gokul Subramanian and Smith, Kaitlin N. and Viszlai, Joshua and Wu, Xin-Chuan and Hardavellas, Nikos and Martonosi, Margaret R. and Chong, Frederic T. },
booktitle = { 2022 IEEE International Symposium on High-Performance Computer Architecture (HPCA) },
title = {{ SupermarQ: A Scalable Quantum Benchmark Suite }},
year = {2022},
volume = {},
ISSN = {},
pages = {587-603},
doi = {10.1109/HPCA53966.2022.00050},
url = {https://doi.ieeecomputersociety.org/10.1109/HPCA53966.2022.00050},
publisher = {IEEE Computer Society},
address = {Los Alamitos, CA, USA},
month =apr}

@article{ibm_melbourne_device,
author = {Moradi, Sasan and Brandner, C. and Spielvogel, Clemens and Krajnc, Denis and Hillmich, Stefan and Wille, Robert and Drexler, Wolfgang and Papp, Laszlo},
year = {2022},
month = {02},
pages = {},
title = {Clinical data classification with noisy intermediate scale quantum computers},
volume = {12},
journal = {Scientific Reports},
doi = {10.1038/s41598-022-05971-9}
}

@article{solid_state,
  author       = {Yao, N. Y. and Jiang, L. and Gorshkov, A. V. and Maurer, P. C. and Giedke, G. and Cirac, J. I. and Lukin, M. D.},
  title        = {Scalable architecture for a room temperature solid-state quantum information processor},
  journal      = {Nature Communications},
  volume       = {3},
  pages        = {800},
  year         = {2012},
  doi          = {10.1038/ncomms1788},
  url          = {https://doi.org/10.1038/ncomms1788}
}

@article{who_is_leading,
  author       = {AbuGhanem, M.},
  title        = {Superconducting quantum computers: who is leading the future?},
  journal      = {EPJ Quantum Technology},
  volume       = {12},
  pages        = {102},
  year         = {2025},
  doi          = {10.1140/epjqt/s40507-025-00405-7},
  url          = {https://doi.org/10.1140/epjqt/s40507-025-00405-7}
}

@INPROCEEDINGS{shor,
  author={Shor, P.W.},
  booktitle={Proceedings 35th Annual Symposium on Foundations of Computer Science}, 
  title={Algorithms for quantum computation: discrete logarithms and factoring}, 
  year={1994},
  volume={},
  number={},
  pages={124-134},
  doi={10.1109/SFCS.1994.365700}}

@inproceedings{grover,
author = {Grover, Lov K.},
title = {A fast quantum mechanical algorithm for database search},
year = {1996},
isbn = {0897917855},
publisher = {Association for Computing Machinery},
address = {New York, NY, USA},
url = {https://doi.org/10.1145/237814.237866},
doi = {10.1145/237814.237866},
booktitle = {Proceedings of the Twenty-Eighth Annual ACM Symposium on Theory of Computing},
pages = {212–219},
numpages = {8},
location = {Philadelphia, Pennsylvania, USA},
series = {STOC '96}
}

@article{comp2arch,
author = {Norbert M. Linke  and Dmitri Maslov  and Martin Roetteler  and Shantanu Debnath  and Caroline Figgatt  and Kevin A. Landsman  and Kenneth Wright  and Christopher Monroe },
title = {Experimental comparison of two quantum computing architectures},
journal = {Proceedings of the National Academy of Sciences},
volume = {114},
number = {13},
pages = {3305-3310},
year = {2017},
doi = {10.1073/pnas.1618020114},
URL = {https://www.pnas.org/doi/abs/10.1073/pnas.1618020114},
eprint = {https://www.pnas.org/doi/pdf/10.1073/pnas.1618020114}}

@article{topological_and_subsystem_codes,
  title = {Topological and Subsystem Codes on Low-Degree Graphs with Flag Qubits},
  author = {Chamberland, Christopher and Zhu, Guanyu and Yoder, Theodore J. and Hertzberg, Jared B. and Cross, Andrew W.},
  journal = {Phys. Rev. X},
  volume = {10},
  issue = {1},
  pages = {011022},
  numpages = {19},
  year = {2020},
  month = {Jan},
  publisher = {American Physical Society},
  doi = {10.1103/PhysRevX.10.011022},
  url = {https://link.aps.org/doi/10.1103/PhysRevX.10.011022}
}

@article{OptQC,
title = {OptQC: An optimized parallel quantum compiler},
journal = {Computer Physics Communications},
volume = {185},
number = {12},
pages = {3307-3316},
year = {2014},
issn = {0010-4655},
doi = {https://doi.org/10.1016/j.cpc.2014.07.022},
url = {https://www.sciencedirect.com/science/article/pii/S0010465514002653},
author = {T. Loke and J.B. Wang and Y.H. Chen}
}

@misc{pandora,
      title={Ultra-Large-Scale Compilation and Manipulation of Quantum Circuits with Pandora}, 
      author={Ioana Moflic and Alexandru Paler},
      year={2025},
      eprint={2508.05608},
      archivePrefix={arXiv},
      primaryClass={quant-ph},
      url={https://arxiv.org/abs/2508.05608}, 
}

@misc{ai-partitioning,
      title={Circuit Partitioning Using Large Language Models for Quantum Compilation and Simulations}, 
      author={Pranav Sinha and Sumit Kumar Jha and Sunny Raj},
      year={2025},
      eprint={2505.07711},
      archivePrefix={arXiv},
      primaryClass={cs.ET},
      url={https://arxiv.org/abs/2505.07711}, 
}

@ARTICLE{modular_compilation_for_DQC,
  author={Ferrari, Davide and Carretta, Stefano and Amoretti, Michele},
  journal={IEEE Transactions on Quantum Engineering}, 
  title={A Modular Quantum Compilation Framework for Distributed Quantum Computing}, 
  year={2023},
  volume={4},
  number={},
  pages={1-13},
  doi={10.1109/TQE.2023.3303935}}

@article{optimised_comp_for_DQC,
author = {Cuomo, Daniele and Caleffi, Marcello and Krsulich, Kevin and Tramonto, Filippo and Agliardi, Gabriele and Prati, Enrico and Cacciapuoti, Angela Sara},
title = {Optimized Compiler for Distributed Quantum Computing},
year = {2023},
issue_date = {June 2023},
publisher = {Association for Computing Machinery},
address = {New York, NY, USA},
volume = {4},
number = {2},
url = {https://doi.org/10.1145/3579367},
doi = {10.1145/3579367},
journal = {ACM Transactions on Quantum Computing},
month = feb,
articleno = {15},
numpages = {29}
}

@ARTICLE{comp_design_for_DQC,
  author={Ferrari, Davide and Cacciapuoti, Angela Sara and Amoretti, Michele and Caleffi, Marcello},
  journal={IEEE Transactions on Quantum Engineering}, 
  title={Compiler Design for Distributed Quantum Computing}, 
  year={2021},
  volume={2},
  number={},
  pages={1-20},
  doi={10.1109/TQE.2021.3053921}}

@inproceedings{gate_commutation,
author = {Itoko, Toshinari and Raymond, Rudy and Imamichi, Takashi and Matsuo, Atsushi and Cross, Andrew W.},
title = {Quantum circuit compilers using gate commutation rules},
year = {2019},
isbn = {9781450360074},
publisher = {Association for Computing Machinery},
address = {New York, NY, USA},
url = {https://doi.org/10.1145/3287624.3287701},
doi = {10.1145/3287624.3287701},
booktitle = {Proceedings of the 24th Asia and South Pacific Design Automation Conference},
pages = {191–196},
numpages = {6},
location = {Tokyo, Japan},
series = {ASPDAC '19}
}

\end{document}